\documentclass[11pt,a4paper]{article}
\usepackage{amssymb}
\usepackage{rotating}
\usepackage{amsfonts}
\usepackage[colorlinks=true, urlcolor=blue, pdfborder={0 0 0}]{hyperref}
\usepackage{amsmath}
\usepackage{amsthm}
\usepackage{caption}
\usepackage{float}
\usepackage{graphicx}
\usepackage{wrapfig}
\usepackage{enumerate}
\usepackage{subcaption}
\usepackage{multicol}
\usepackage{srcltx}
\usepackage{array}
\newtheorem{theorem}{Theorem}[section]
\newtheorem{proposition}{Proposition}[section]
\newtheorem{lemma}{Lemma}[section]

\newtheorem{remark}{Remark}
\newcommand{\wh}{\widehat}
\newcommand{\ov}{\overline}
\newcommand{\wt}{\widetilde}

\newcommand\cE{{\cal E}}

\newcommand\cF{{\cal F}}

\newcommand\cN{{\cal N}}

\newcommand\e{{\varepsilon}}

\def\bbn{{\mathbb N}}

\def\bbr{{\mathbb R}}

\def\text#1{\hbox{#1}}
\def\proof{{\noindent \bf Proof. }}

\def\endproof{\mbox{\ $\qed$}}

\def\E{{\bf E}}

\def\P{{\bf P}}
\def\C{{\bf C}}
\def\L{{\bf L}}

\def\B{{\bf B}}

\def\H{{\bf H}}

\def\r{{\bf r}}
\def\x{{\bf x}}
\def\a{{\bf a}}
\def\b{{\bf b}}

\def\v{{\bf v}}

\def\Chi{{\bf 1}}
\def\d{\mathrm{d}}
\def\build #1_#2{\mathrel{\mathop{\kern 0pt #1}\limits_\zs{#2}}}
\newcommand{\zs}[1]{{\mathchoice{#1}{#1}{\lower.25ex\hbox{$\scriptstyle#1$}}
{\lower0.25ex\hbox{$\scriptscriptstyle#1$}}}}

\numberwithin{equation}{section}

\def\proof{{\noindent \bf Proof. }}
\def\endproof{\mbox{\ $\qed$}}

\usepackage[top=1.5in, bottom=1.5in, left=1.1in, right=1in]{geometry}

\begin{document}

\title{Approximate hedging with constant proportional transaction costs in financial markets with jumps}

\author{ Thai Nguyen\thanks {University of Ulm and University of Economics Hochiminh city; Email: thaibopy@ueh.edu.vn; thai.nguyen@uni-ulm.de
} \and Serguei Pergamenschchikov\thanks{Laboratoire de Math\'{e}matiques Rapha\"{e}l Salem, UMR 6085 CNRS-Universit\'{e} de Rouen, France 
and 
International Laboratory SSP \& QF, Tomsk State University, Russia,
 email: serge.pergamenchtchikov@univ-rouen.fr.
}}
\maketitle

\begin{abstract}
We study the problem of option replication under constant proportional transaction costs in models where stochastic volatility and jumps are combined to capture the market's important features. Assuming some mild condition on the jump size distribution we show that transaction costs can be approximately compensated by applying the Leland adjusting volatility principle and the asymptotic property of the hedging error due to discrete readjustments is characterized. In particular, the jump risk can be approximately eliminated and the results established in continuous diffusion models are recovered. The study also confirms that for the case of constant trading cost rate, the approximate results established by Kabanov and Safarian \cite{Kab-Saf1} and by Pergamenschikov \cite{Per} are still valid in jump-diffusion models with deterministic volatility using the classical Leland parameter in \cite{Leland}.
\end{abstract}


\vspace{2mm}

\section{Introduction}
Many suggested mathematical models for stock prices have been trying to capture important markets features, e.g. \emph{leptokurtic feature, volatility clustering effect, implied volatility smile}. These market properties are tractable in stochastic volatility models. However, diffusion-based stochastic volatility models assume that the market volatility can fluctuate autonomously but can not change suddenly and as a result, they could not take into account sudden and unpredictable market changes. Hence, in more realistic settings (extensions of the famous Black-Scholes framework), the continuity assumption of stock price should be relaxed to account for sudden market shocks due to good or bad market news. These sudden events can arrive according to a Poisson process. The change in the asset price right after the market receives a (good/bad) news can be described by a jump size and between two consecutive jump times, the asset price follows a geometric Brownian motion as in the classical Black-Scholes models. Such a combination is called a jump-diffusion model. As shown in \cite{Kou}, jump-diffusion models not only fit the data better than the classical geometric Brownian
motion, but also well reproduce the leptokurtic feature of return distributions. Moreover, \cite{Rama} argues that the presence of jumps in the asset price can be recognized as the presence of participants in the option market. See \cite{Rama,Rug,Kou} and the
references therein for detailed discussions.

Note that in complete diffusion models, options can be completely replicated using the delta strategy which is adjusted continuously. However, it is not the case for models with jumps. In fact, the jump risk can not be released completely even under continuous time strategies and the only way to hedge perfectly a Call option against the jump risk is to buy and hold the underlying asset. In other words, in the presence of jumps, the conception of replication does not indicate a right framework for risk management and hedging like in diffusion-based complete market models where the Black-Scholes theory plays a central role. 

The situation becomes more challenging if one takes transaction costs into account. Such a consideration is realistic and has been attractive to researchers for the last two decades. Intuitively, in the presence of transaction costs and/or jumps in the asset price, the option is more risky and should be evaluated at higher price than that in the absence of these risks. However, a more expensive option price would suggest an increase in its volatility values. This is the essential intuition behind the Leland algorithm. In particular, to compensate trading costs in the absence of jumps, Leland \cite{Leland} proposed a modified version of the well-known Black-Scholes PDE where the volatility is artificially increased as 
\begin{equation}
\wh{\sigma}^2=\sigma^2+\sigma \kappa n^{1/2-\alpha}\sqrt{8/\pi},
\label{eq:Intro.1}
\end{equation}
where $n$ is the number of revisions and $\kappa n^{-\alpha}, \, 0\le \alpha\le 1/2$ is the cost rate. He claimed that the option payoff $h(S_1)$ (maturity $T=1$) can be asymptotically replicated by $V^n_1$, the terminal portfolio value of the discrete delta strategy, as the hedging time distances become small for $\alpha =0$ (constant rate) or $\alpha=1/2$. Kabanov and Safarian \cite{Kab-Saf1} showed later that the Leland statement for constant transaction cost is not mathematically correct and the hedging error in fact converges to a non-zero limit $\min(S_1,K)-J(S_1)$ as the portfolio is frequently revised. Here $J(S_1)$ is the limit of cumulative costs. The rate of convergence and the asymptotic distribution of the corrected replication error was then investigated in \cite{Per}. In particular, the latter paper showed that the sequence 
\begin{equation}
n^{1/4}(V^n_1-h(S_1)-\min(S_1,K)+\kappa J(S_1))
\label{eq:Perg}
\end{equation} weakly converges to a mixed Gaussian variable as $n\to\infty$.
This result has initiated many further studies in different directions: general payoffs with non-uniform readjustments \cite{Lepinette08,Lepinette10,Darses}, local volatility by \cite{LepinetteTran}, trading costs based on the traded number of asset by \cite{LepinetteELie}. Recently, \cite{Nguyen} has showed that the increasing volatility principle is still helpful for controlling losses caused by trading costs which are proportional to the trading volume in stochastic volatility frameworks with a simpler form for the adjusted volatility. Furthermore, the latter paper also pointed out a connection to asset hedging in high frequency market, where the form of bid-ask price could be an essential factor for deciding laws of trading costs. We refer the reader to the papers mentioned above and the references therein for more detailed discussions.

\vspace{2mm}

The problem of hedging under transaction costs has been developed in various directions.  In the context of small proportional transaction costs, \cite{Grannan96,ahn1998,Fukasawa11,Fukasawa2014} apply a sequence of {\it stopping times} as rebalancing dates to study the hedging error. As discussed in \cite{cai2016}, non of these strategies outperform the others. 
The latter paper \cite{cai2016} suggests a class of continuous control strategies that lead to finite transaction costs and provides a condition under which the option payoff is asymptotically replicated. However, as mentioned above continuous trading strategies cannot be used in practice. In general, trading activities are only possible in a specific period of day. Therefore, discrete time strategies with periodically rebalancing dates are of highly important relevance.

In this paper, we contribute to the field of approximate hedging under transaction costs using Leland's algorithm by allowing for jumps. To our best knowledge, we are the first to study the approximate hedging problem under trading costs for models with jumps, using a discrete time strategy resulting from the Black-Schole pricing PDE with a modified volatility. The aim of the present note is to build a bridge from the existing results in continuous diffusion models studied by e.g. \cite{Kab-Saf1,Per,Lepinette08,Lepinette10,Darses,Nguyen,Nguyen2} to discontinuous models where jumps are allowed in the asset price and/or volatility. In fact, we try to capture not only the dependence structure (modelled by stochastic volatility) but also short term behaviors of the stock price due to sudden market changes\footnote{This fact partially explains why jump-diffusion models are, in general, considered as a good choice, especially in short-term situations.}. As stochastic volatility models well complement models with jumps \cite{Kou}, this combination leads to a realistic and general model of financial markets.

As mentioned above, the jump risk in the hedging problem is challenging to handle and can not be released completely even in simple framework e.g. jump-diffusion models where continuous adjustments are possible. The contribution of this paper is twofold. First, we show that impact of jumps can be partially negligible under some mild condition on jump sizes. In fact, we prove that the asymptotic distribution of the hedging error is independent of jumps and consistent with the result of pure diffusion models established in \cite{Nguyen}. The same is true when jumps are allowed in both the asset price and its volatility. Such general frameworks provide some possibility to explain large movements in volatility which are usually observed during crisis periods \cite{Eraker03,Eraker04}. As the second contribution, we show that the Kabanov-Safarian-Pergamenshchikov results in \cite{Kab-Saf1,Per} also hold for jump-diffusion settings. We finally remark that extensions of these results to the case of general convex payoffs that satisfy some decaying assumption \cite{Lepinette08,Lepinette10,Darses} are feasible.

\vspace{2mm}
The remainder of this paper is organized as follows. We shortly present the key idea behind the Leland algorithm in Section \ref{AD}. Section \ref{Mol} is devoted to formulate the model and present our main results. General stochastic volatility models with jumps are discussed in Section \ref{GeneralJSV}. We consider a special case when volatility is deterministic in Section \ref{Constant} and present a numerical example in Section \ref{Num}. Proof of main results are reported in Section \ref{Proof}. Some useful Lemmas can be found in the Appendix.

\section{Leland's algorithm with a modified volatility}\label{AD}
To explain the key idea in the Leland's algorithm we assume that the stock price is given by $\d S_t =\sigma S_t \d W_t$ in the hedging interval $[0,1]$ and the interest rate is zero so that $S$ is a martingale under the risk-neutral measure. Under the presence of proportional transaction costs, it was proposed by \cite{Leland} then generalized by \cite{Kab-Saf1,Kab-Saf2} that the volatility should be adjusted as in equation \eqref{eq:Intro.1} in order to create an artificial increase in the option price $C(t,S_t)$ to compensate possible trading fees. This form is inspired from the observation that the trading cost $\kappa_n S_\zs{t_i}|C_x(t_\zs{i},S_\zs{t_{i}})-C_x(t_\zs{i-1},S_\zs{t_{i-1}})|$ in the interval of time $[t_\zs{i-1},t_i]$ can be approximated by 
\begin{equation}
\kappa_nS_\zs{t_{i-1}}C_{xx}(t_\zs{i-1},S_\zs{t_{i-1}})|\Delta S_\zs{t_i}|\approx \kappa_n\sigma S^2_\zs{t_{i-1}}C_{xx}(t_\zs{i-1},S_\zs{t_{i-1}}) \E |\Delta W_\zs{t_i}|.
\label{eq:rev.1}
\end{equation}
For simplicity, we assume that the portfolio is revised at uniform grid $t_i=i/n, i=1,\dots, n$ of the option life interval $[0,1]$. Taking into account that $\E |\Delta W_\zs{t_i}/(\Delta t_i)^{1/2}|=\sqrt{2/\pi}$ one approximates the last term in \eqref{eq:rev.1} by $\kappa_n\sigma \sqrt{2/\pi} (\Delta t_i)^{1/2}S^2_\zs{t_{i-1}}C_{xx}(t_\zs{i-1},S_\zs{t_{i-1}})$, which is the cost paid for the portfolio readjustment in $[t_\zs{i-1},t_i]$. Hence, by the standard argument of the Black-Scholes theory, the option price inclusive of trading cost should satisfy
$$
C_t(t_\zs{i-1},S_\zs{t_{i-1}}) \Delta t_i+\frac{1}{2}\sigma^2 S_\zs{t_{i-1}}^2 C_{xx}(t_\zs{i-1},S_\zs{t_{i-1}})\Delta t_i+\kappa_n\sigma \sqrt{2/\pi} (\Delta t_i)^{1/2} S_\zs{t_{i-1}}^2 C_{xx}(t_\zs{i-1},S_\zs{t_{i-1}}) =0.
$$
Since $\Delta t_i=1/n$, one deduces that 
$$
C_t(t_\zs{i-1},S_\zs{t_{i-1}}) +\frac{1}{2}\left(\sigma^2 +\kappa_n\sigma \sqrt{n 8/\pi}\right ) S_\zs{t_{i-1}}^2 C_{xx}(t_\zs{i-1},S_\zs{t_{i-1}}) =0,
$$
which implies that the option price inclusive trading cost should be evaluated by the following modified-volatility version of the Black-Scholes PDE
\begin{equation}
\wh{C}_t(t,x) +\frac{1}{2}\wh{\sigma}^2 x^2 \wh{C}_{xx}(t,x) =0,\quad \wh{C}(1,x)=\max(x-K,0),
\label{eq:rev.2}
\end{equation}
where the adjusted volatility $\wh{\sigma}$ is defined by \eqref{eq:Intro.1}.
\begin{figure}[h]
\begin{subfigure}[b]{0.5\textwidth}
\includegraphics[width=\columnwidth,height=6cm]{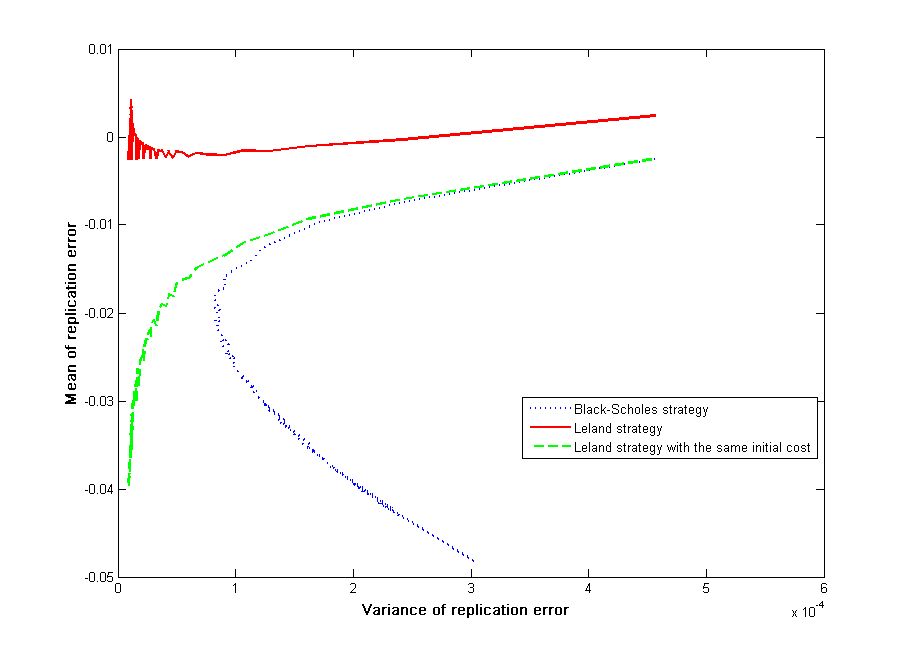}%
\caption{\scriptsize Mean-variance}%
\label{fg.Ch_1.BSvsLeland.1}%
\end{subfigure}%
\begin{subfigure}[b]{0.5\textwidth}
  \includegraphics[width=\linewidth, height=6cm]{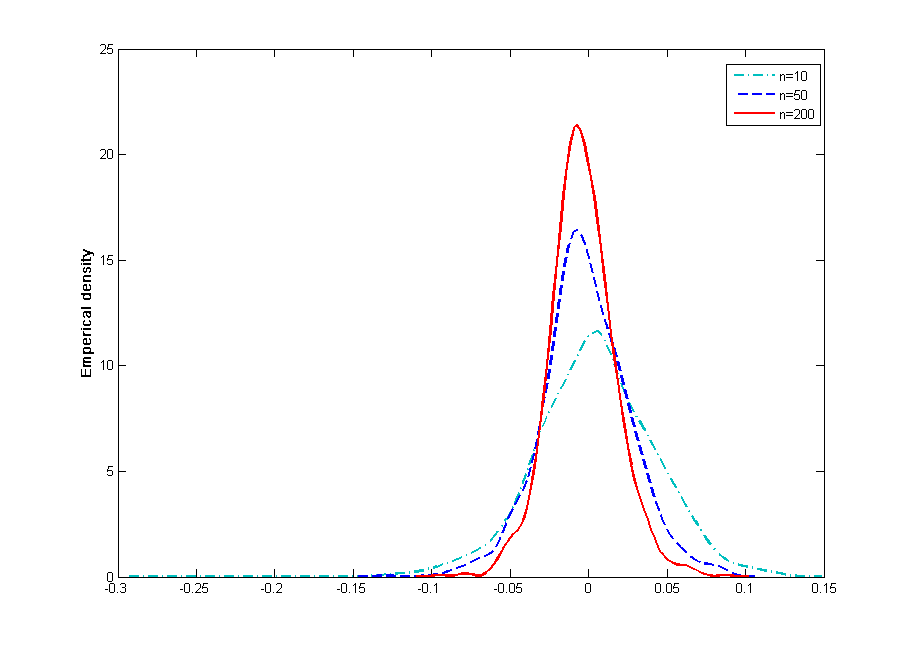}
\caption{\scriptsize Empirical density }
\label{fig.Ch_1.empr.1}
 \end{subfigure}
\caption{\small Mean-variance of Leland strategy and its normalized corrected hedging error}\label{fig.Density.01} 
\end{figure}

The simulation results for the profit/loss $V^n_1-h(S_1)$ in Figure \ref{fg.Ch_1.BSvsLeland.1} suggest that in the presence of transaction costs, Leland strategy outperforms \footnote{ A strategy is better than the others if it offers the lowest risk for a given level of returns. In the literature, the expectation replication error, also known as the expectation profit and loss, seems to be a good return measure while the variance of replication error is the most popular risk measure. A highly-tolerant hedger would prefer a position with high mean replication error and a high risk, whereas a highly risk-averse hedger would prefer a point with a low risk and low mean replication.} the Black-Scholes discrete delta strategy even when they starts with the same initial cost. The empirical distribution and density of the normalized corrected hedging error in the Pergamenschikov's theorem \cite{Per} (see eq.\eqref{eq:Perg}) are shown in Figure \ref{fig.Ch_1.empr.1}. It turns out that the corrected hedging empirically converges to a centered variable that likely behaves as a normal distribution but with an asymmetric tail.

\vspace{2mm}
As seen above, the Leland method is important for practical purposes since it requires only a small change in parameter of the well-known Black-Scholes framework which is widely used in practice. However, when volatility is random, say $\sigma=\sigma(y_t)$ where $y_t$ is another random process, the strategy is no longer available from the Cauchy problem \eqref{eq:rev.2}. In fact, pricing and hedging in such stochastic volatility (SV) contexts are intrinsically different from the Leland model and even in the absence of trading costs, the option price strongly depends on volatility level and future information of the volatility process $y$. To see this issue, let us assume that $y$ is a Markov process. By the iterative property of expectation, the option price exclusive of trading costs is then given by 
\begin{equation}
C(t,x,y)=\E[C_\zs{BS}(t,x;K,\ov{\sigma})|y_t=y],
\label{eq:}
\end{equation}
where $K$ is the strike price, $C_\zs{BS}$ is the Black-Scholes option price functional and $\ov{\sigma}$ is the averaged volatility defined by
$$
\ov{\sigma}^2=\frac{1}{1-t}\int_t^1 \sigma^2(y_s) \d s.
$$
It means that the option price is the average of Black-Scholes price on all possible future trajectories of the volatility process $y$, which in reality cannot be observed directly. Thus, the option price and hence hedging strategy in SV models are very complicated and usually studied via asymptotic analysis.

Now let us now turn our attention to the presence of transaction costs. The above discussion has emphasized that the well-known form \eqref{eq:Intro.1} for adjusted volatility in Leland's algorithm is no longer helpful from a practical point of view. Furthermore, even for local volatility models in which $\sigma$ only depends on time and the spot price $S$, it is technically difficult to show the existence of a solution to \eqref{eq:rev.2} since the differential operator is not uniformly parabolic. In addition, estimates for derivatives of the option price, which are essential for approximation analysis, are not easy to achieve, see e.g. \cite{Lepinette08}.

Fortunately, a deep study on the approximation of the hedging error shows that the limit of the replication error does not strongly depend on the form of adjusted volatility but only on the last term $\kappa_n \sqrt{n}$ whenever the latter product diverges to infinity. This important observation means that a simpler form $\varrho\kappa_n \sqrt{n}$ with some constant $\varrho$, can be used to obtain the same asymptotic property for the hedging error. This modification is fully investigated for SV models in \cite{Nguyen,Nguyen14}. Note that the option price and hedging strategy can be easily obtained if the new form is applied for the Cauchy problem \eqref{eq:rev.2} while it is impossible in practice for the classical one \eqref{eq:Intro.1}. In this paper we show that this suggestion is still useful even when jumps in the asset price and/or in stochastic volatility are taken into account.

We conclude the section by mentioning some important features of hedging in jumps models. First, jump risks cannot be covered by simply using the classical delta strategy even with a continuous-time adjustment policy. Second, if jumps are allowed in the stock price then one should distinguish two types of hedging errors: one is due to the market incompleteness concerning jumps and the other one is due to the discrete nature of the hedging portfolio. These two types of hedging errors have different behaviors. The literature of discrete hedging with jumps is vast and we only mention to \cite{TanVol,TanBro} for recent achievements. However, none of these mentioned papers discuss about trading costs.

\section{Model with jumps and main results}\label{Mol}
\subsection{The market model}
Let $\left( \Omega ,\mathcal{F}_{1},\left( \mathcal{F}_\zs{t}\right) _{0\leq
t\leq 1},\mathbf{P}\right) $ be~the standard filtered probability space with
two standard independent $\left( \mathcal{F}_\zs{t}\right) _{0\leq t\leq 1}$
adapted Wiener processes $(W_\zs{t}^{\left( 1\right) })$ and $(W_\zs{t}^{\left(
2\right) })$ taking values in $\mathbb{R}$. Consider a financial market
that consists of one non-risky asset set as the \emph{num\'eraire} and the risky
one (e.g. \emph{stock}) $S_\zs{t}$ defined as
\begin{equation}
\label{sec.Mol.1}
\left\{ 
\begin{array}{l}
\d S_\zs{t}=S_\zs{t^{-}} \left(b_t \d t+\sigma \left(y_\zs{t}\right)\d W_\zs{t}^{\left( 1\right)}+\d \zeta_\zs{t}\right), \\ 
\d y_\zs{t}=\alpha_1\left( t,y_\zs{t}\right) \d t+\alpha_2\left( t,y_\zs{t}\right)\d W^{(2)}_\zs{t},
\end{array}
\right.
\end{equation}%
where $
S_\zs{t^{-}}=\lim_{s\uparrow t}S_{s}$ and 
  the jump L\'evy process is defined as
\begin{equation}
\label{jumps-1}
\zeta_\zs{t}=x*(J-\nu)_\zs{t},
\end{equation}
which is independent of the Wiener processes  $(W_\zs{t}^{\left( 1\right) })$ and $(W_\zs{t}^{\left(2\right) })$.
Here  $J(\d t,\d x)$ is the jump random measure of $\zeta$ having the deterministic compensator 
$\nu(\d t,\d x)=\d t\,\Pi(\d x)$, where $\Pi(\cdot)$ is the L\'evy measure defined for any Borel set $A\subseteq\bbr_\zs{*}=\bbr\setminus\{0\}$
as
\begin{equation}
\label{levy-measure}
\Pi(A)=\E\,\sum_\zs{0\le s\le 1}\Chi_\zs{\{\Delta \zeta_\zs{s}\in A\}}
\quad\mbox{and}\quad
 \Delta \zeta_\zs{s}=\zeta_\zs{s}-\zeta_\zs{s-}.
\end{equation}
In order for $S_\zs{t}$ to be positive
 we assume that $\Pi(]-\infty,-1])=0$. Moreover, we assume that
\begin{equation}
\label{levy-measure_condition}
\Pi(x^{2})
=
\int_\zs{\bbr_*}z^{2}\Pi(\d z)<\infty.
\end{equation}
In the sequel we use the notation $\Pi(f)=\int_\zs{\bbr_*}f(z)\Pi(\d z)$.
We recall that the symbol $*$ means the stochastic integral, i.e.
for any $t>0$ and any function $h:[0,t]\times\bbr\to\bbr$, 
$$
h*(J-\nu)_\zs{t}=
\int^{t}_\zs{0}\,
\int_{\bbr_\zs{*}}\,h(u,x)\,\d \wt{J}(\d u\,,\, \d x)
\,,\quad
\wt{J}(\d t\,,\,\d x)=(J-\nu)(\d t\,,\d x).
$$
We assume that the coefficients $\alpha_i, i=1,2$ are locally Lipschitz and linearly growth functions, which provides the existence of the unique strong solution $y$ to the second equation \cite{Fried}. Moreover, we assume that  the processes $(W^{(1)}_\zs{t})_\zs{0\le t\le 1}$, $(W^{(2)}_\zs{t})_\zs{0\le t\le 1}$ and $(\zeta_\zs{t})_\zs{0\le t\le 1}$ are independent. Note that, if the volatility function $\sigma(\cdot)$ is constant, we obtain a L\'evy financial market, which is very popular in the finance, see e.g. \cite{kabanov2016ruin} and the references therein.  We assume further that $\Delta \zeta_\zs{t}>-1$,
i.e. $\Pi(-\infty\,,\,-1])=0$. So, in this case using the Dol\'eans-Dade exponent we can represent the price process as
\begin{equation}\label{Sec.Mol.2++}
S_\zs{t}=S_\zs{0}\exp\left\{\int_{0}^{t} \check{b}_\zs{u}\, \d u+\int_{0}^{t}\sigma \left( y_{s}\right) \d W^{(1)}_{s}
+ 
\check{\zeta}_\zs{t}
\right\},
\end{equation}
where
$\check{\zeta}_\zs{t}=\ln(1+x)*(J-\nu)_\zs{t}$ and $\check{b}_\zs{t}=b_\zs{t}-\sigma^{2}(y_\zs{t})/2+\Pi(\ln(1+x)-x)$. 

Note that the L\'evy measure defined in \eqref{levy-measure} may be infinite, i.e. $\Pi(\bbr_\zs{*})=+\infty$.
If $\theta=\Pi(\bbr_\zs{*})<\infty$,
 we can represent the jump process in \eqref{sec.Mol.1} as
\begin{equation}\label{Sec.Mol.Pois_mod} 
 \zeta_\zs{t}=\sum_{j=1}^{N_\zs{t}}\xi _{j}
 -
 \theta
 \E\xi_\zs{1}\,
 t,
\end{equation} 
 where $(N_\zs{t})_\zs{t\ge 0}$ is
a Poisson process  with intensity parameter $\theta$
 and $(\xi_j)_\zs{j\ge1}$ are i.i.d. 
 random variables taking values in $(-1,+\infty)$. 
We assume further that the Poisson process $N_t$ and the jumps sizes $(\xi_j)_\zs{j\ge1}$ are independent. In this case, the stock price can be expressed as  
\begin{equation}\label{Sec.Mol.2}
S_\zs{t}=S_{0}\exp \left\{ \int_{0}^{t} \check{b}_\zs{u}\, \d u+\int_{0}^{t}\sigma \left( y_{s}\right) \d W^{(1)}_{s}
+\sum_{j=1}^{N_\zs{t}}\ln
\left( 1+\xi _{j}\right) \right\},
\end{equation}%
where
$\check{b}_\zs{t}=b_\zs{t}-\sigma^{2}(y_\zs{t})/2-\theta\E\,\xi_\zs{1}$.


\begin{remark}
The drift $b_t$ plays no role in approximation. In fact, our asymptotic results are valid for any $b_t$ satisfying $\sup_{0\le t\le 1}\E \vert b_t\vert<\infty$.
\end{remark}
%
\begin{remark}\label{Re.1}
 In this paper we will not discuss about the problem of change of measure and the jump risk but we accept the free-risk assumption of asset dynamics as the starting point. Clearly, a jump-diffusion setting leads to an incomplete market, which is also an important feature of stochastic volatility settings. Hence, there are many ways to choose the pricing measure via Girsanov's technique. Such a procedure makes an essential change not only on the diffusion but also on the jump part of the asset dynamics \cite{Rama,Merton}. In \cite{Kou}, an expectation equilibrium argument is used to obtain a simple transform from the original physical probability to a risk-neutral probability for which many assets (bonds, stocks, derivatives on stocks) can be simultaneously priced in the same  framework. 
\end{remark}

\subsection{Assumptions and examples}
The following condition on the jump sizes is accepted in our consideration:

{\em
\noindent $(\C_1)$ The L\'evy measure of jump sizes satisfies
\begin{equation*}\label{Sec.Mol.20}
\Pi(\bbr_\zs{*})<\infty 
\quad \mbox{and} \quad \int_\zs{(-1,0]}(1+z)^{-1}{\Pi(\d z)}<\infty.
\end{equation*}
}
The first integrability condition is nothing than the condition of finite variance for the jump size distribution. The second one is equivalent to $\E(1+\xi)^{-1}<\infty$. These conditions are not too strong and automatically fulfilled in Merton jump-diffusion models \cite{Merton} where the jump size distribution is assumed to be log-normal. In \cite{Kou} within an equilibrium-based setting, log-exponential distributions are suggested to achieve convenient features in an analytical calculation. Note that this family of jump size distributions also satisfies Condition $(\C_1)$. 

Let us turn our attention to volatility assumptions. Following \cite{Nguyen} we assume that the volatility process satisfies the integrable condition.

\noindent $(\C_2)$
{\em $\sigma$ is a twice continuously differentiable function satisfying 
$$
0<\sigma_\zs\min \le \inf_\zs{y\in\bbr}\sigma(y)\quad \mbox{and}\quad \sup_\zs{0\le t\le 1}\E\, \max\{\sigma(y_t), \vert\sigma'(y_t)\vert\}<\infty.
$$
}
In fact, Condition $(\C_2)$ is fulfilled for almost widely used stochastic volatility models, see \cite{Nguyen} for more discussions.

\vspace{2mm}

\begin{remark}\label{Re.2}
 Remark that in the present setting, the combination of stochastic volatility and jumps means that the asset price is not a L\'evy process but a semi-martingale. As mentioned in \cite{Nguyen}, finite moments of the asset process in SV models are not guaranteed in general, see \cite{And-Pit,Lions-Mus}. This crucial feature prevent us from making an $L^2$-based approximation as in deterministic volatility models \cite{Kab-Saf1,Lepinette08,Lepinette10}. Therefore, for the approximation procedures below we follow the approach in \cite{Nguyen,Nguyen2} which is based on a truncation technique. The convergence results obtained in what follows are only achieved in probability.
\end{remark}

We conclude this subsection with some well-known stochastic volatility models with jumps, see \cite{Rama} and Section \ref{GeneralJSV} for more examples.

\vspace{2mm}
\noindent {\bf Bates model}: The Bates model is a jump-diffusion stochastic volatility model obtained by adding proportional log-normal jumps to the Heston stochastic volatility model:
\begin{equation}
\d S_t=S_t(a_\zs{0} \d t+\sqrt{y_t} \d W_t^S+\d Z_t);\quad \d y_t=a(m-y_t)\d t+b\sqrt{y_t}\d W_t^y,
\end{equation}
where $W^S,W^y$ are Brownian motions with correlation $\rho$ and $Z$ is a compound Poisson process with intensity $\theta$ and log-normal distribution. Condition $(\C_1)$ is clearly verified since jumps follow the log-normal law. 
The Bates model exhibits some nice properties from a practical point of view. Firstly, the characteristic function of the log-price is available in a closed-form, which is important for pricing purposes. Secondly, the implied volatility pattern for long term and short term options can be adjusted separately \cite{Rama}.

\vspace{2mm}
\noindent{\bf Ornstein-Uhlenbeck stochastic volatility models}: It is possible to introduce a jump component in both price and volatility processes. Such models are suggested by Barndorff-Nielsen and Shephard  to take  leverage effects into account:
\begin{equation}
S_\zs{t}=S_\zs{0} \,e^{X_\zs{t}}\,,
\quad  \d X_\zs{t}=(a_\zs{0}+a_\zs{1} \sigma_t^2)\d t+\sigma_t \d W_t+\rho 
\d Z_\zs{t}\,,\quad \d\sigma_t^2=-\theta \sigma_t^2\d t+\d Z_t.
\end{equation}
If $\rho=0$ the volatility moves with jumps but the price process has continuous paths. The case $\rho\neq 0$, representing a strong correlation between volatility and price, is more flexible but computation is now challenging. Remark that in this case $\sigma$ is not the ''true'' volatility as the returns are also affected by changes of the L\'evy process $Z_t$. If jumps still follow a log-normal law then Condition $(\C_1)$ is fulfilled. 

\subsection{Approximate hedging with transaction costs: main result}\label{Result}
In this section, we study the problem of discrete hedging under proportional transaction costs using the increasing volatility principle as in Leland's algorithm following the setting in \cite{Nguyen}. More precisely, we suppose that for each successful transaction, traders are charged by a cost that is proportional to the trading volume with the cost coefficient $\kappa$. Here $\kappa$ is a positive constant defined by market moderators.  Let us suppose that the investor plans to revise his portfolio at dates $(t_{i})$ defined by
\begin{equation}\label{Sec.Mol.5}
t_{i}=g\left( i/n\right) ,\quad 
g(t)=1-\left( 1-t\right)^{\mu }\,,\quad
1\le \mu<2\,,
\end{equation}
where $n$ is the number of revisions. The parameter $\mu$ is used to control the rate of convergence of the replication error. The bigger value of $\mu$ the more frequently the portfolio is revised. Clearly one gets the well-known uniform readjustment for $\mu=1$. 

To compensate transaction costs caused by hedging activities, the option seller is suggested to follow the
Leland strategy defined by the piecewise process 
\begin{equation}\label{Sec.Mol.14}
\gamma _\zs{t}^{n}=\sum_{i=1}^{n}\hat{C}_{x}\left( t_{i-1},S_{t_{i-1}}\right) 
\mathbf{1}_{(t_{i-1},t_{i}]}\left( t\right),
\end{equation}
 where $\wh{C}$ is the solution to the following adjusted-volatility Black-Scholes PDE
\begin{equation}
C_\zs{t}(t,x)+\frac{1}{2}\wh{\sigma}_t^2 x^2 C_\zs{xx}(t,x)=0,\quad 0\le t<1, \quad C(1,x)=h(x):=(x-K)_\zs{+}\,.
\label{Sec:nocost.1}
\end{equation}
Here the adjusted volatility $\wh{\sigma}^2$ is given by
\begin{equation}\label{Sec.Mol.6}
\wh{\sigma}^2(t)=\varrho\sqrt{n f'(t)}\quad \mbox{with}\quad f(t)=g^{-1}(t)=1-(1-t)^{1/\mu}.
\end{equation}
The motivation of this simple form is discussed in Section \ref{AD}, see more in \cite{Nguyen}. We remark that when volatility is deterministic the classical modified volatility form \eqref{eq:Intro.1} can be used. This special case will be treated in Section \ref{Constant}. Now, using strategy $\gamma_t^n$ requires a cumulative trading volume measured in dollar value which is given by 
$$
\Gamma_n=\sum^{n}_\zs{i=1}\,S_\zs{t_\zs{i}}\vert \gamma_\zs{t_\zs{i}}^{n}-\gamma
_\zs{t_\zs{i-1}}^{n}\vert\,.
$$
By It\^o's lemma one represents the payoff as 
\begin{align*}
h(S_1)&=\wh{C}(0,S_0)+ \int_0^1 \wh{C}_\zs{x}(t,S_\zs{t^{-}})\d S_t
\\[2mm]
&
+\int_0^1\left(\wh{C}_\zs{t}(t,S_\zs{t})
+ \frac{1}{2}\sigma^2(y_t)S^2_\zs{t} \wh{C}_\zs{xx}(t,S_\zs{t})\right)\d t 
+\sum_\zs{0\le t\le 1}
\,\B(t,S_\zs{t^{-}},\Delta S_\zs{t}/S_\zs{t^{-}})\,,
\end{align*}
where
$\B(t,x,z)
=\hat{C}\left( t,x(z+1)\right)
-\hat{C}\left( t,x\right)-zx\hat{C}_\zs{x}\left( t,x\right)$
and
 $\Delta S_\zs{t}=S_\zs{t}-S_\zs{t^{-}}$ is the jump size of the stock price at time $t$. 
The above sum of jumps can be represented as 
\begin{equation}\label{Sec.Mol.14-1-1-0}
I_\zs{3,n}=\int_{0}^{1}\int_\zs{\bbr_\zs{*}}\B\left( t,S_\zs{t^{-}},z\right){J}(\d t\,,\, \d z)\,.
\end{equation}
Assuming that the initial capital (option price) is given by $V^n_0=\wh{C}(0,S_0)$ and using \eqref{Sec:nocost.1}, one represents \footnote{Note that for Lebesgue  and It\^o's integrals one can replace $S_\zs{t^{-}}$ by $S_\zs{t}$.} the hedging error as
 \begin{align}\label{Sec.Mol.14-1}
V^{n}_\zs{1}-h(S_\zs{1})=\frac{1}{2}I_\zs{1,n}+I_\zs{2,n}-I_\zs{3,n}-\kappa\Gamma_n,
\end{align}
where 
$I_\zs{1,n}=\int_\zs{0}^{1}\,
\left(\wh{\sigma}_\zs{t}^{2}-\sigma ^{2}(
y_\zs{t})\right)\, S_\zs{t}^{2}\wh{C}_\zs{xx}( t,S_\zs{t}) \d t
$ and $I_\zs{2,n}=\int_\zs{0}^{1}\left(\gamma_\zs{t^{-}}^{n}-\wh{C}_\zs{x}(t,S_\zs{t^{-}})\right) \d S_\zs{t}$.

The goal now is study asymptotic property of the replication error $V^{n}_\zs{1}-h(S_\zs{1})$. To describe asymptotic properties, let us introduce the following functions
\begin{equation}\label{Sec.Mol.23-1}
\v(\lambda,x)=\frac{\ln(x/K)}{\sqrt{\lambda}}+\frac{\sqrt{\lambda}}{2},\quad q(\lambda,x)=\frac{\ln(x/K)}{2\lambda}-\frac{1}{4} \quad \mbox{and}
\quad
\wt{\varphi}(\lambda,x )= \varphi\left(\v(\lambda,x)\right),
\end{equation}
where $\varphi$ is the standard normal density function.
As shown below, the rate of convergence of our approximation will be controlled by the parameter $\beta$ defined by
\begin{equation}
 \frac{1}{4}\le\beta:=\frac{\mu}{2(\mu+1)}<\frac{1}{3}, \quad\mbox{for}\quad 1\leq \mu <2.
\end{equation}

Before stating our main result, let us emphasize that using an enlarged volatility which diverges to infinity implies that the asymptotic property of the hedging error strongly depends on trading times near by the maturity. But remember that jumps are rare events and hence, possible jumps near by the maturity can be omitted with a very small probability. Therefore, jumps in such contexts do not much affect the asymptotic property of the hedging error as the hedging revision is taken more frequently. This implies that increasing volatility is still helpful for models with jumps. The theorems below are generalizations of the achievement of continuous stochastic volatility models in \cite{Nguyen}. 

\begin{theorem}\label{Th.sec:Mol.1} Under conditions $(\C_1)-(\C_2)$, the sequence of
$$n^{\beta }\left( V^n_1-h(S_1)-\min \left(
{S}_{1},K\right)+\kappa \Gamma\left( {S}_{1},y_{1},\varrho\right) \right)$$ converges to a centered-mixed Gaussian variable as $n$ tends to infinity, where the positive function $\Gamma$ is the limit of trading volume defined as
\begin{equation}
\Gamma\left( x,y,\varrho \right)= x\int_\zs{0}^{+\infty }
\,
\lambda ^{-1/2}\wt{\varphi}(\lambda,x ) 
\,\E\,
\left|
\sigma(y){\varrho}^{-1}
Z+q(\lambda,x)
\right| 
\,
\d\lambda,
\label{eq:0}
\end{equation}
in which $Z$ is a standard normal variable independent of $S_{1},y_1$.
\end{theorem}
The term $q(\lambda,x)$ in the limit of transaction costs can be removed using a modified Leland strategy (the so-called L\'epinette's strategy) defined by
\begin{equation}\label{Sec.Mol.14-2}
\bar{\gamma} _\zs{t}^{n}=\sum_{i=1}^{n}\left(\hat{C}_{x}(t_{i-1},S_{t_{i-1}})-\int_0^{t_\zs{i-1}}\hat{C}_{xt}( t,S_\zs{t})\d t
\right)\mathbf{1}_{(t_{i-1},t_{i}]}(t).
\end{equation}
\begin{theorem}\label{Th.sec:Mol.2}
Suppose that L\'epinette's strategy is used for option replication. Then, under Conditions $(\C_1)-(\C_2)$ the sequence of $n^{\beta }\left( \bar{V}^n_1-h(S_1)-\eta\min \left(
{S}_{1},K\right) \right)$ converges to a centered-mixed Gaussian variable as $n$ tends to infinity, where $\eta=1-\kappa\sigma(y_1){\varrho}^{-1}\sqrt{8/\pi}$.
\end{theorem}

\subsection{Super-hedging and price reduction}\label{sec:Superhedge}
We first emphasize that it is
impossible to obtain a \emph{non-trivial} perfect hedge with the presence of
transaction costs even in constant volatility models. To
cover 
completely the option return, the seller can take the 
\emph{buy-and-hold} strategy, but this makes the option price too expensive. Cvitani\'{c} and Karatzas \cite{Cvitanic} show
that the \emph{buy-and-hold strategy} is the unique choice if one wishes to successfully replicate the option and $S_\zs{0}$ is
the super-replication price under the presence of transaction costs. 
As proved in \cite{Nguyen}, a suitable choice of $\varrho$ can lead to super-replication in the limit.

\begin{proposition}
Let Conditions $(\C_1)-(\C_2)$ hold and $\sigma$ be a twice continuously differentiable and bounded function. Then there exists $\varrho_*>0$ such that 
$\lim_{n\to\infty} V^n_1\ge h(S_1)$ for any $\varrho \ge \varrho_*$. This property is true for both Leland's strategy and L\'epinette's strategy.
\end{proposition}
The superhedging cost is too high from the buyer's point of view though it indeed gives the seller a successful hedge with probability one in the limit.
More practically, one can ask how much initial capital can be reduced by accepting a shortfall probability in replication objective. From the Black-Scholes formula one observes that both strategies $\gamma^n_t$ and $\bar{\gamma}^n_t$ approach to the buy-and-hold one as $n\to\infty$. In \cite{Per,Nguyen} a simple method is suggested to lower the option price following the quantile hedging spirit. Let us adapt  the main idea in these works for the present setting. Since $S_\zs{1^{-}}=S_\zs{1}$ almost surely, we define
\begin{equation}\label{sec:Ap.5}
\delta_\zs{\varepsilon }=\inf \left\{ a>0:\Upsilon( a) \geq 1-\varepsilon
\right\},  
\end{equation}
where $\Upsilon(a) =\P\left((1- 
\kappa) \min (S_\zs{1},K)
>( 1-a) S_\zs{0}\right)$. The quantity $\delta_\zs{\varepsilon }$ is called quantile price of the option at level $\varepsilon$ and the difference $(1-\delta_\zs{\varepsilon }) S_0$ is the reduction amount of the option price (the initial cost for quantile hedging), see \cite{Follmer,Novikov,Per,Baran,Bratyk,Barski} for more discussions.
 Clearly, the smaller value of $\delta_\zs{\varepsilon }$ is, the cheaper the option is. We show that the option price is significantly reduced, compared with power functions of the parameter $\varepsilon $.
\begin{proposition}\label{Pr.sec:Ap.1}
Let $\delta_\zs{\varepsilon }$ be the Leland price defined by 
\eqref{sec:Ap.5} and assume that the jump sizes are almost surely non-negative, i.e. 
\begin{equation}\label{sec:Ap.5-0-0}
\xi_j\ge 0,\quad \mbox{a.s.}\quad\forall j\in\bbn,
\end{equation}
and $
\sigma_\zs{\max}=
\sup_\zs{y\in\bbr}\sigma (y)<\infty\,.
$
Then, for any 
$r>0$, 
\begin{equation}\label{sec:Ap.5-0}
\lim_\zs{\varepsilon \rightarrow 0}
({1-\delta_\zs{\varepsilon }}){\varepsilon^{-r}}=+\infty\,.
\end{equation}
\end{proposition}
\proof
Observe that $0<\delta_\zs{\varepsilon }\leq 1$ and $\delta_\zs{\varepsilon }$ 
tends to $1$ as $\varepsilon \rightarrow 0$. Set
$b=1-\kappa$. Then for sufficiently small $\varepsilon$ such that 
$\delta_\varepsilon >a> 1-bK/S_\zs{0}$ one has
\begin{align*}
1-\varepsilon 
&> \P( \min(S_\zs{1},K)>(1-a) S_\zs{0}) =1-\P(S_\zs{1}/S_\zs{0}\leq ( 1-a) ) .
\end{align*}
Therefore, 
\begin{equation}\label{sec:Ap.5-1}
  \varepsilon 
< \P\left(S_\zs{1}/S_\zs{0}\le (1- a)(1-\kappa)^{-1}\right)=\P\left(P_1(\xi){\cal E}_1(y)\le z_a \right),
 \end{equation}
where $z_a=(1- a)(1-\kappa)^{-1}e^{\lambda \E\xi_1}$ and
\begin{equation}\label{sec:Ap.5-1-0}
\cE_\zs{t}(\sigma)=\exp\left\{\int_\zs{0}^{t}\,\sigma(y_s)\d W^{(1)}_\zs{s}
-\frac{1}{2}\int_\zs{0}^{t}\sigma^2(y_s)\d s\right\} \quad \mbox{and} \quad 
P_t(\xi)= \prod_\zs{j=1}^{N_t}(1+\xi_j).
\end{equation}
By \eqref{sec:Ap.5-0-0}, $P_t(\xi)\ge 1$ for all $t\in[0,1]$, which implies that the probability in the right side of \eqref{sec:Ap.5-1} is bounded by 
$P\left({\cal E}_1(y)\le z_a\right)$. Therefore, $  \varepsilon 
< \P\left({\cal E}_1(y)\le z_a \right).$
At this point, the conclusion exactly follows from Proposition 4.2 in \cite{Nguyen} and the proof is completed.\endproof

\section{General stochastic volatility models with jumps}\label{GeneralJSV}
\subsection{Introduction}
Stochastic volatility with jumps in price (SVJP) models have been very popular in the option pricing literature as they provide flexibilities to capture important features of returns distribution. However, empirical studies \cite{Eraker03,Eraker04} show that they do not well reflex large movements in volatility assets during periods of market stress such as those in 1987, 1997, 2008. In other words, SVJP models are misspecified for such purposes. These studies also suggest that it would be more reasonable to add an extra component into the volatility dynamics so that this new factor allows volatility to rapidly increase. Note that such expected effect can not be generated by only using jumps in returns (as in jump-diffusion models) or diffusive stochastic volatility. In fact, jumps in returns can only create large movements but they do not have future impact on returns volatility. On the other hand, diffusive stochastic volatility driven by a Brownian motion only generates small increase via sequences of small normal increments. Many empirical studies \cite{Eraker03,Eraker04} show that incorporating jumps in stochastic volatility can successfully capture rapid changes in volatility.

It is important to note that introducing jumps in volatility does not imply an elimination of jumps in returns. Although jumps both in returns and volatility are rare, each of them plays an important part in generating crash-like movements. In crisis periods, jumps in returns and in volatility are more important factor than the diffusive stochastic volatility in producing large increases. We refer the reader to \cite{Eraker03,Eraker04} for more influential discussions about financial evidence for the use of jumps in volatility. 

In this section, we study the problem of option replication under transaction costs in a general SV model which allows for jumps in both the asset price and volatility. This is clearly a generalization of the setting in Section \ref{Mol}. We show that jumps in volatility can be also ignored as those in the asset price, i.e. the results obtained in Section \ref{Mol} can be recovered.

\subsection{Specifications of SV models with jumps}
Assume that under the objective probability measure, the dynamics of stock prices $S$ are assumed to be given by 
\begin{equation}\label{sec.Extension.0-1}
\d S_\zs{t}=S_\zs{t^{-}} \left(b_t(y_\zs{t})\d t+\sigma (y_\zs{t})\d W_\zs{t}^{(1)} + \d \zeta_\zs{t}^S\right),\quad
\d y_\zs{t}=\alpha_1\left( t,y_\zs{t}\right) \d t+\alpha_2\left( t,y_\zs{t}\right) \d W^{(2)}_\zs{t}+\d \zeta_\zs{t}^y.
\end{equation}
Here, $\zeta_\zs{t}^S=\sum_\zs{j=1}^{N_t^S} \xi_j^S$ and $\zeta_\zs{t}^y=\sum_\zs{j=1}^{N_t^y} \xi_j^y$ are two compound Poisson processes. For a general setting, two Poisson processes $N_t^r$ and two sequence of jump sizes $(\xi_j^r)$, $r\in\{S,y\}$ can be correlated. Let us give some possible specifications for the jump components.

\begin{enumerate}

	\item[{\bf (i)}] \noindent{\it Stochastic volatility model (SV)}: This corresponds to the case when there is no jump in both the asset price and volatility, i.e. $\zeta_\zs{t}^S=\zeta_\zs{t}^y=0,\forall t$. This basic SV model has been widely investigated in the literature. The problem of approximate hedging under proportional transaction costs is studied in \cite{Nguyen,Nguyen2}. Roughly speaking, adding some extra component generated by a diffusion to the returns distribution of a classic Black-Scholes setting gives a SV model. 
	
	\vspace{1mm}

\item[{\bf (ii)}] \noindent{\it Stochastic volatility with jumps in volatility (SVJV)}: By allowing jumps in volatility process $y$ one can obtain an extension of SV models, i.e. $\zeta_\zs{t}^S=0, \forall t$ but $ \zeta_\zs{t}^y\neq 0$. In such cases, option pricing implications are in fact inherited from SV models.

\vspace{1mm}

\item[{\bf (iii)}] \noindent{\it Stochastic volatility with jumps in price (SVJP)}: Assume now that $\zeta_\zs{t}^S\neq 0$ but $ \zeta_\zs{t}^y=0$. This case is studied in Section \ref{Mol}.

\vspace{1mm}
\item[{\bf (iv)}] 
\noindent{\it Stochastic volatility with common jumps in price and volatility (SVCJ)}: Suppose that both the asset price and its volatility in a SV model are influenced by the same extra random factor modeled by a compound Poisson process. In other words, jumps in the asset price and in volatility are driven by the same compound Poisson process $\zeta_\zs{t}^S=\zeta_\zs{t}^y$.

\vspace{1mm}
\item[{\bf (v)}] 
\noindent{\it Stochastic volatility with state-dependent and correlated jumps (SVJJ)}: This is the most general case for the present setting \eqref{sec.Extension.0-1}.
\end{enumerate}
\subsection{Option replication with transaction costs in general SVJJ models}
In this subsection we study the problem of option replication presented in Section \ref{Mol} for general SVJJ models \eqref{sec.Extension.0-1}. We show that in the same hedging policy as in SVJP defined in Section \ref{Mol}, jump effects can be ignored in asset as well as in volatility. First, let us recall from Section \ref{Mol} that the hedging error takes the form
$$V^{n}_\zs{1}-h(S_\zs{1})=\frac{1}{2}I_\zs{1,n}+I_\zs{2,n}-I_\zs{3,n}-\kappa \Gamma_n,$$
where $I_\zs{i,n},\,i=1,2,3$ and $\Gamma_n$ are defined as in \eqref{Sec.Mol.14-1}. The following conditions on volatility dynamics are needed in this section.

\vspace{2mm}

{\em
\noindent $(\C_3)$ The coefficient functions $\alpha_i, i=1,2$ are linearly bounded and locally Lipschizt. 
 }
\vspace{1mm} 
 
Condition $(\C_3)$ implies that $\sup_{0\le t\le 1}\E \, y_\zs{t}^2<\infty$, which is necessary for approximation procedure.

\begin{theorem}\label{Th.Extension.1} Under conditions $( \C_1)-(\C_2)-(\C_3)$, the limit results in Theorems \ref{Th.sec:Mol.1} and \ref{Th.sec:Mol.2} still hold.
\end{theorem}

\section{Deterministic volatility models with jumps}\label{Constant}
In this section we consider a simpler model where the asset volatility is a constant, i.e. $\sigma(y)=\sigma=\mbox{constant}>0,\, \forall y$. In the absence of jumps in the asset price, it is well-known that using the classical adjusted volatility
\begin{equation}
\wh{\sigma}^2_t=\sigma^2+ \varrho_0\sqrt{n f'(t)},\quad \varrho_0=\kappa \sigma \sqrt{8/\pi}
\label{eq:const.1}
\end{equation}
leads to a non-zero discrepancy between asymptotic portfolio value and the option payoff. In particular, \cite{Kab-Saf1} proved that $V^n_1$ converges in probability to $$h(S_1)+\min(S_1,K)-\kappa J(S_1,\varrho_0),$$ where $\Gamma(x,\varrho)$ defined by \eqref{eq:0} with $\sigma(y)=\sigma$. It was then proved in \cite{Per} that asymptotic distribution of the normalized corrected hedging error is a mixed Gaussian. In particular, for Leland's strategy and uniform revisions, $$n^{1/4}(V^n_1-h(S_1)-\min(S_1,K)+\kappa J(S_1,\varrho))$$ converges weakly to a centered mixed Gaussian variable for any $\varrho>0$. In order to remove the corrector in Pergamenshchikov's result, the paper \cite{Darses} applied the L\'epinette strategy with $\wh{\sigma}^2$ defined in \eqref{eq:const.1}. In fact, it is proved that for general European options with a payoff function $h$ verifying a power decay property, $n^{\beta}(V^n_1-h(S_1))$ converges to a mixed Gaussian variable. Note that these mentioned results are obtained in the a diffusive models setting for the asset price. 

When jumps are present in the asset price, the approximation in these works should be reconsidered heavily and seemingly this desired extension is far away from being obvious in their approach.

One advantage of our method is that possible jumps in the asset price can be ignored with a very small probability. In fact, as claimed below the same result is true for jump-diffusion models.

\begin{theorem}\label{Th.Const.1}
Assume that the asset dynamics is given by 
$$
\d S_\zs{t}=S_\zs{t^{-}} \left(b_t\, \d t+\sigma \d W_\zs{t}+\d\bigg(\sum_{j=1}^{N_t} \xi_j \bigg)\right), 
$$ 
where $N_t$ is a Poisson process with intensity $\theta$ and $(\xi_j)_\zs{j\ge 1}$ is an i.i.d sequence of random variable whose common distribution satisfies Condition $(\C_1)$, $W$ is a Brownian motion independent of the compound Poisson process $\sum_{j=1}^{N_t} \xi_j$  and $b$ is a bounded continuous deterministic function. Suppose further that the adjusted volatility $\wh{\sigma}$ is defined by \eqref{eq:const.1} or by the simple form \eqref{Sec.Mol.6}. Then, for Leland's strategy, the normalized hedging error $$n^\beta\big(V^n_1-h(S_1)-\min(S_1,K)+\kappa J(S_1,\varrho_0)\big)$$ converges weakly to a centered mixed Gaussian variable. This is still true for any $\varrho>0$ in the place of $\varrho_0$ and hence the result in \cite{Per} is recovered.

If the L\'epinette strategy with $\wh{\sigma}^2$ defined by \eqref{eq:const.1} is applied one gets an asymptotic complete replication, i.e. $n^\beta(\ov{V}^n_1-h(S_1))$ converges weakly to a centered mixed Gaussian variable. 
\end{theorem}

\begin{remark}
Theorem \ref{Th.Const.1} can be extended to the case where the volatility is a bounded and smooth deterministic function with a convex general payoff holding a power decaying condition like in \cite{Darses}.
\end{remark}

\section{Numerical example}\label{Num}
\begin{figure}[h]
\begin{subfigure}[b]{0.5\textwidth}

  \includegraphics[width=\columnwidth,height=6cm]{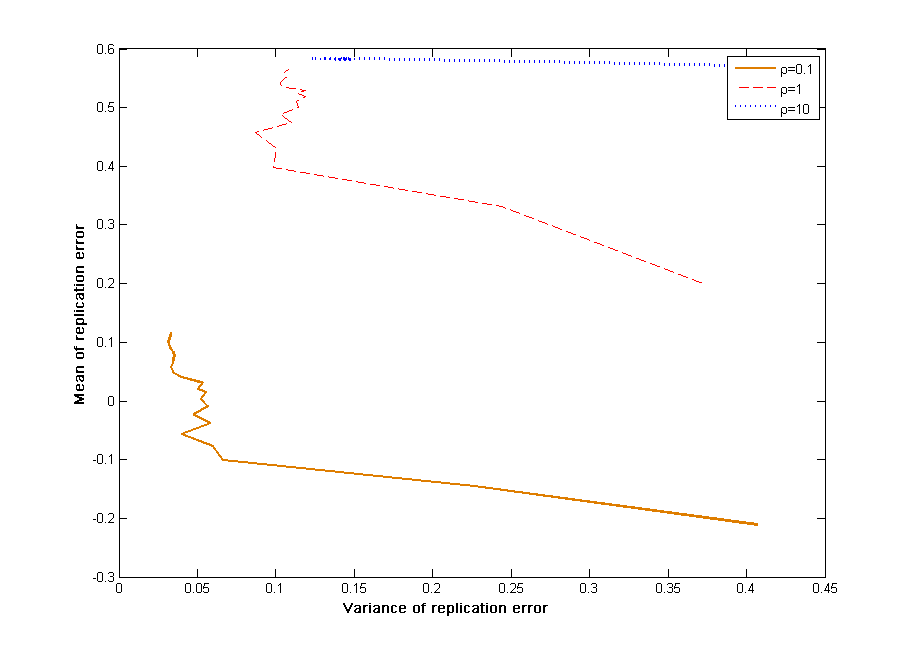}
\caption{\small Variance and mean of Leland strategy}%
\label{fig.mv.jump}
\end{subfigure}%
\begin{subfigure}[b]{0.5\textwidth}
\includegraphics[width=\columnwidth,height=6cm]{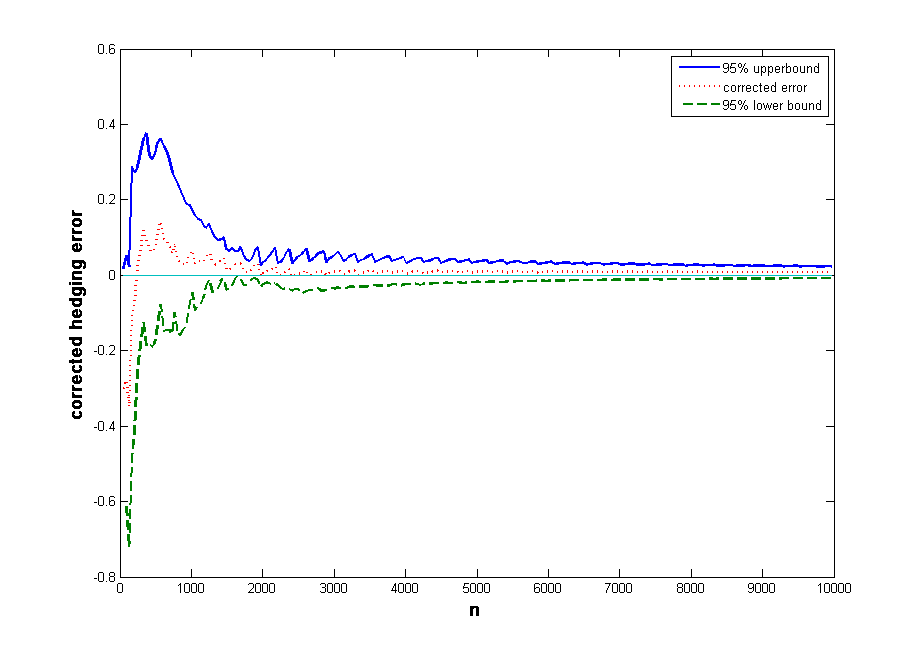}
\caption{\small Corrected hedging error with $\kappa=0.001$}%
\label{fig.Error.1}
 \end{subfigure}
\caption{\small Mean-variance of Leland strategy and its normalized corrected hedging error}\label{fig.Ch_5.Density.2} 
\end{figure}

We present in this section a numerical example using Matlab 2012b. In particular, we assume that the asset price follow the jump-diffusion model with stochastic volatility driven by an Orstein-Uhlenbeck process $\d y_t =(a-y_t)\d t +b \d W_t$ and volatility function $\sigma(y)=y_0 e^{y}+\sigma_{min}$. This can be considered as a Hull-White's model with jumps allowed in the asset price. The jump size distribution is $N(0,0.2)$ and the jump rate is $\theta=3$. The other parameters are chosen as $S_0=K=1$, $a=-1, b=0.2$, $\sigma_\zs{min}=1$, $y_0=2$, $\varrho=\sqrt{8/\pi}$. Figure \ref{fig.mv.jump} depicts the numerical hedging error in the mean-variance space. 
%
The convergence to zero of the corrected hedging error $V^n_1-h(S_1)-\min(S_1,K)-\kappa J(S_1,y_1,\varrho)$ is somehow slow, see Figure \ref{fig.Error.1}, where the mean value of the corrector is 0.2465.


\begin{figure}[H]%
\centering\includegraphics[width=0.8\linewidth, height=7cm]{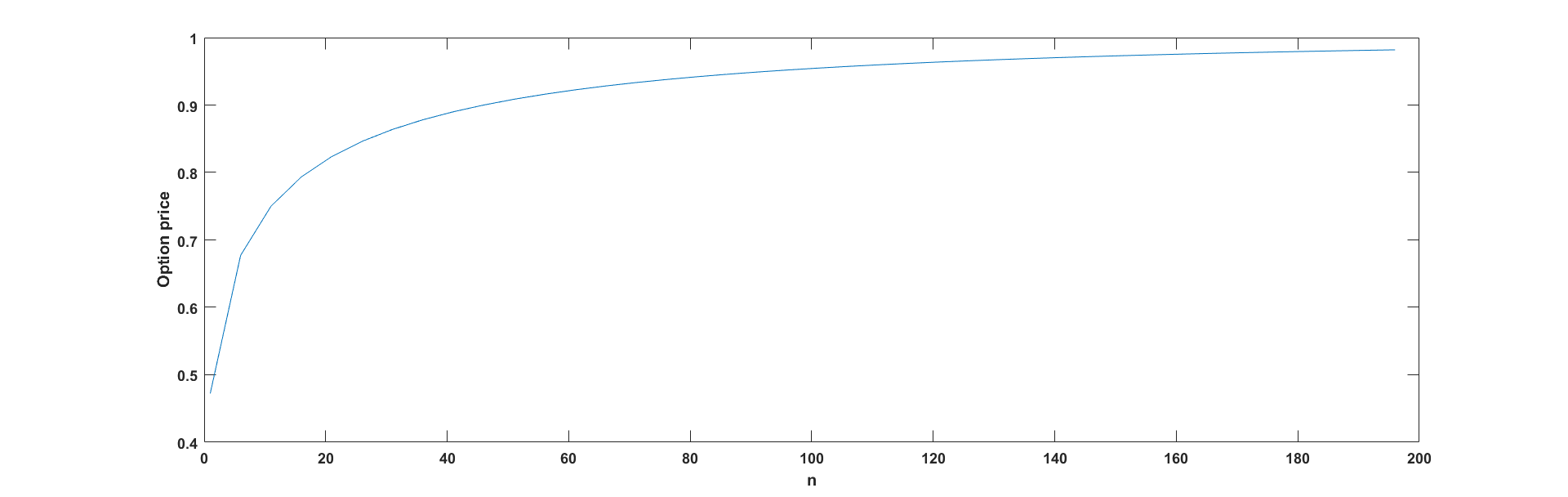}%
\caption{Option price at time $t=0$.}%
\label{Fig.Optionprice}
\end{figure}

\begin{figure}[h]
\begin{subfigure}[b]{0.5\textwidth}
  \includegraphics[width=\linewidth, height=5cm]{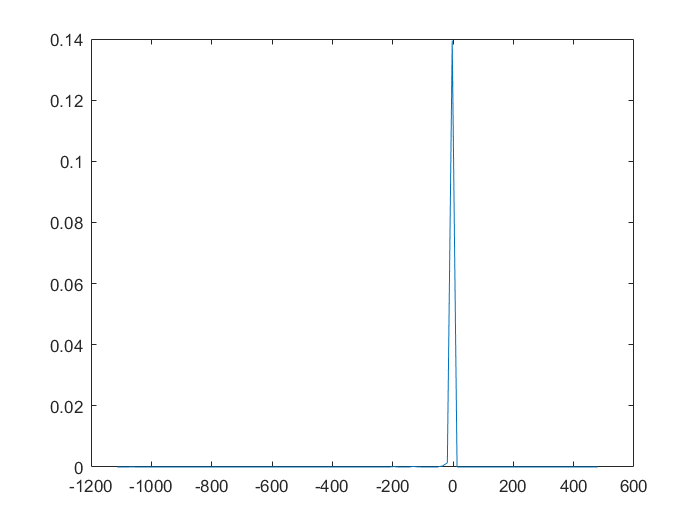}
\caption{\small $\varrho=0.01$, $n=1000$}\label{fig. Lepinettedenrho001}
\end{subfigure}%
\begin{subfigure}[b]{0.5\textwidth}
  \includegraphics[width=\linewidth, height=5cm]{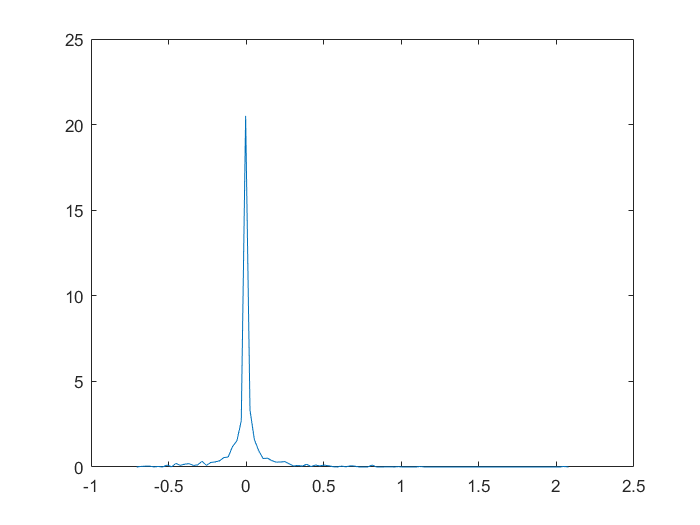}
\caption{\small $\varrho=5$,$n=1000$}\label{fig. Lepinettedenrho5}
 \end{subfigure}
\caption{\small Empirical density of the normalized corrected hedging error L\'epinette strategy}\label{fig.Lepinettedensity} 
\end{figure}

\begin{figure}[h]%
\centering\includegraphics[width=0.8\linewidth, height=8cm]{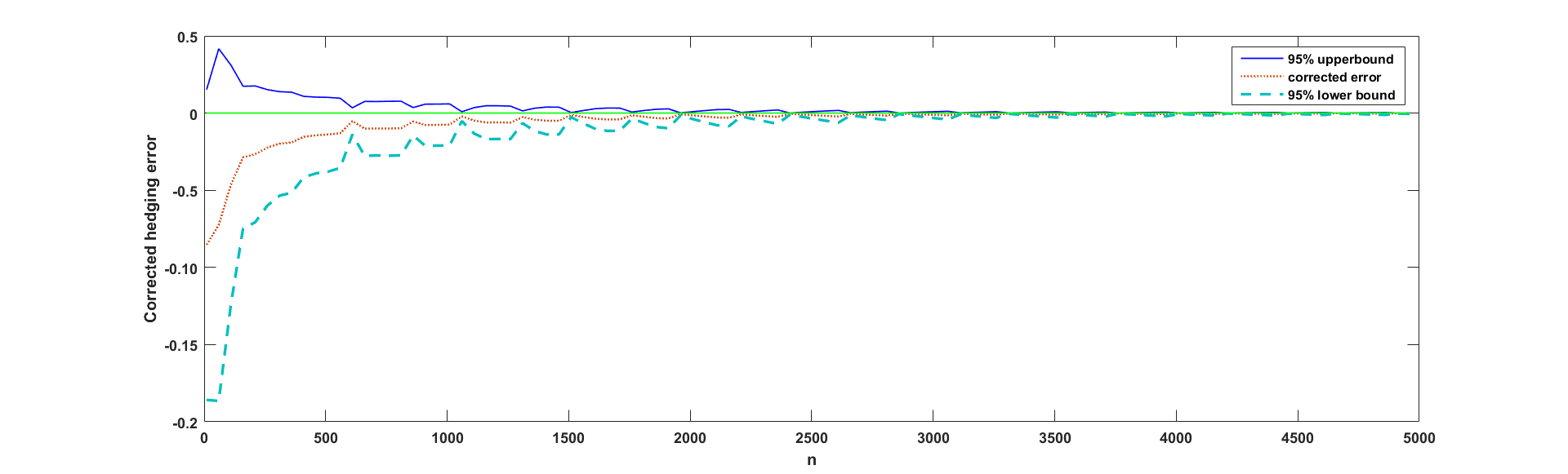}%
\caption{Corrected hedging error of L\'epinette strategy}%
\label{Fig.Lepinetteerror}
\end{figure}

It can be observed in Figure \ref{Fig.Optionprice} that the option price at time zero converges to the buy-and-hold superhedging price $S_0=1$. This numerically confirms that the quantile hedging approach discussed in Section \ref{sec:Superhedge} could be used to reduce the price, see e.g. \cite{Per,Nguyen} for more detail.
The empirical density of the normalized corrected hedging error is reported in Figure \ref{fig.Ch_5.Density.2}. Not surprisingly, we observe that the hedging empirically converges to a centered variable that likely behaves as a  normal distribution but with a significantly asymmetric tail. Indeed, the tail of the limiting distribution is more complex than that corresponding to the case without jumps illustrated in Figure \ref{fig.Density.01}.
\begin{figure}[h]
\begin{subfigure}[b]{0.5\textwidth}
  \includegraphics[width=\linewidth, height=5cm]{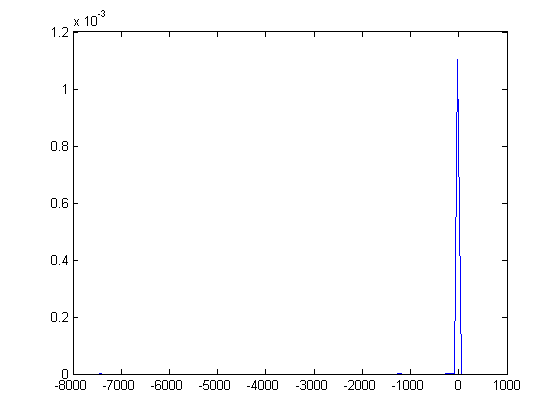}
\caption{\small $\varrho=0.01$, $n=1000$}
\end{subfigure}%
\begin{subfigure}[b]{0.5\textwidth}
  \includegraphics[width=\linewidth, height=5cm]{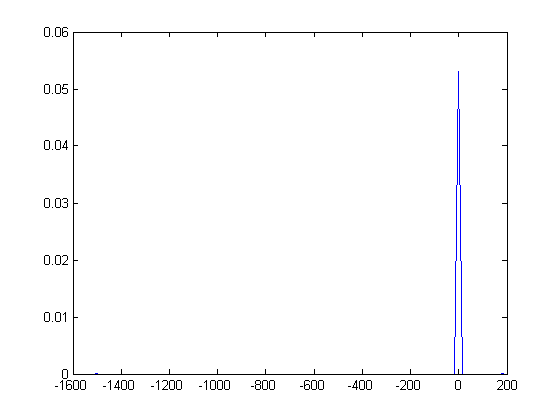}
\caption{\small $\varrho=5$,$n=1000$}
 \end{subfigure}
\caption{\small Empirical density of the normalized corrected hedging error}\label{fig.Ch_5.Density.2} 
\end{figure}
In Figure \ref{fig.Lepinettedensity}, we also observe that the normalized hedging error for L\'epinette strategy defined in \eqref{Sec.Mol.14-2} empirically converges to a centered variable that behaves as a normal distribution. Compared to Leland strategy, the limiting variance is less asymmetric. In addition, as observed in Figure \ref{fig. Lepinettedenrho5}, increasing $\varrho$ leads to a more normal-liked limiting distribution. This is due to the fact that the use of a correcting term (in integral form) in \eqref{Sec.Mol.14-2} has reduced the limiting variance significantly. By the same reason, Figure \ref{Fig.Lepinetteerror} shows a numerically improved convergence for the hedging error, compared to the Leland strategy.

\section{Proof of Main Theorems}\label{Proof}
As usual, the main results established in three
Sections \ref{Mol}, \ref{GeneralJSV} and \ref{Constant} are direct consequences of some specific types of limit theorem for martingales that we are searching for. For this aim, we construct a special approximation procedure following the one in \cite{Nguyen}. Our main attempt is to prove that the jump terms appearing in the approximation can be neglected at the desired rate $n^\beta$. For convenience, we recall in the first subsection the preliminary setup and refer to \cite{Nguyen} for the motivation.
\subsection{Preliminary}\label{sec:Bat}
Define
$m_\zs{1}=n-\left[n\left( {l^{*}}/{\lambda_\zs{0}}
\right)^{2/( \mu +1) }\right]$ and
$  m_\zs{2}=n-\left[n\left({l_{*}}/{\lambda_\zs{0}}
\right)^{2/(\mu +1)}\right],$
where $[x]$ stands for the integer part of a number $x$ and $
l_\zs{*}=\ln^{-3}n,\ l^{*}=\ln^{3}n
$. Below we focus on the subsequence $(t_j)$  of trading times 
and the corresponding sequence
$\left(\lambda_\zs{j}\right)_\zs{m_\zs{1}\le j\le m_\zs{2}}$
 defined as 
\begin{equation}\label{sec:Bat.3}
t_\zs{j}=1-(1-j/n) ^{\mu }
\quad\mbox{and}\quad
\lambda_\zs{j}=\int_{t_j}^1 \wh{\sigma}^2_u \d u=
\lambda_\zs{0}(1-t_\zs{j})^\frac{1}{4\beta},\quad
\lambda_\zs{0}=\frac{4\beta\sqrt{n}}{\sqrt{\mu}}\,.
\end{equation}
Note that
$\left(t_\zs{j}\right)$ 
is an increasing sequence
 with values in $\left[t^{\ast},t_{\ast}\right]$, where
$t_{\ast}=1-( l_\zs{\ast}/\lambda_\zs{0}) ^{4\beta}$ and
$t^{\ast}=1-(l^{*}/\lambda_\zs{0})^{4\beta}$, whereas $\left(\lambda_\zs{j}\right)$ is decreasing in $[l_*,l^*]$.
Therefore, in the sequel we make use the notation $\Delta t_j= t_j-t_{j-1}$ whereas $\Delta \lambda_j= \lambda_{j-1}-\lambda_{j}$, for $m_1\le j\le m_2$ to avoid the negative sign in discrete sums. 

\begin{figure}[h]
\begin{center}
\includegraphics[width=0.8\columnwidth,height=7cm]{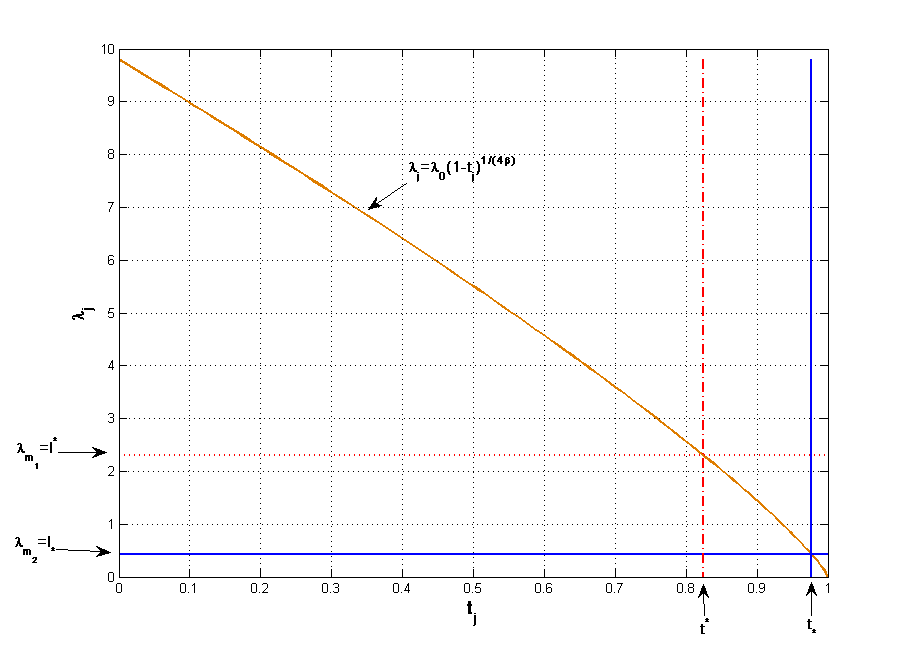}%
\caption{Two sequences $(\lambda_j)$ and $(t_j)$}%
\label{fig.Twosequence}%
\end{center}
\end{figure}

Below, It\^o integrals will be discretized throughout the following  sequences of independent normal random variables
\begin{equation}\label{sec:Tool.12-1}
{Z}_\zs{1,j}=\frac{W_\zs{t_\zs{j}}^{(1)}-W_\zs{t_\zs{j-1}}^{(1)}}
{\sqrt{t_\zs{j}-t_\zs{j-1}}}\qquad \mbox{and}\qquad
{Z}_\zs{2,j}=
\frac{W_\zs{t_\zs{j}}^{(2)}-W_\zs{t_\zs{j-1}}^{(2)}}{\sqrt{t_\zs{j}-t_\zs{j-1}}}.
\end{equation}
We set 
\begin{equation}\label{sec:Bat.3-1}
{p}(\lambda,x,y)
=\frac{\varrho}{\sigma(y)}\,
\left(
\frac{\ln (x/K)}{2\lambda}
-
\frac{1}{4}
\right)
\end{equation}
and write for short ${p}_{j-1}={p}(\lambda_\zs{j-1},S_\zs{t_\zs{j-1}},y_\zs{t_\zs{j-1}})$. This reduced notation is also frequently applied for functions appearing in the approximation procedure. With the sequence of revision times 
$(t_\zs{j})$ in hand, we consider the centered sequences
\begin{equation}
\begin{cases}
 {Z}_\zs{3,j}=
\vert {Z}_\zs{1,j}+{p}_{j-1}\vert-
\E\,\left(\vert {Z}_\zs{1,j}+{p}_{j-1}\vert\,\vert\,\cF_\zs{j-1}\right),\label{sec.Tool.12}\\[1mm]
{Z}_\zs{4,j}=
\vert {Z}_\zs{1,j}\vert-
\E\,\left(\vert {Z}_\zs{1,j}\vert\,\vert\,\cF_\zs{j-1}\right)=\vert {Z}_\zs{1,j}\vert-\sqrt{2/\pi}.
\end{cases}
\end{equation}
The sequences $( {Z}_\zs{3,j})$ and $( {Z}_\zs{4,j})$ will serve in finding the Doob decomposition of the considered terms. To represent the limit of transaction costs, we introduce the functions
\begin{equation}\label{sec:Bat.3-0}
\begin{cases}
G(a) =\E\,
\left( \left\vert Z+a\right\vert \right)
=2\varphi ( a) +a\left(2\Phi ( a) -1\right),\\[1mm]
\Lambda (a)=\E\,
\left( \vert Z+a\vert -\E\, \vert Z+a\vert \right)^2= 1+a^2-G^2(a),
\end{cases}
\end{equation}
for $a\in\bbr$ and $Z\sim \cN(0,1)$.
We also write $o( n^{-r})$ for generic sequences of random variables 
$(X_\zs{n})_\zs{n\ge 1}$ satisfying $
\P-\lim_\zs{n\rightarrow\infty }n^{r}\,X_\zs{n}=0$ while the notation $X_n=O( n^{-r})$ means that $n^{r}\,X_\zs{n}$ is bounded in probability. For approximation analysis, we will make use of the functions
\begin{equation}\label{sec:Bat.4-1}
\phi(\lambda,x)= \exp\left\{-\frac{x^2}{2\lambda}-\frac{\lambda}{8}\right\}, \quad \wh{\phi}(\lambda,x)=\phi(\lambda,\eta(x))\quad 
\quad
\mbox{with}
\quad\eta(x)=\vert\ln(x/K)\vert.
\end{equation}

\subsection{Stopping time and technical condition }


We first emphasize that in bounded volatility settings, it is possible to carry out an asymptotic analysis based on $L^2$ estimates as in the previous works \cite{Darses,Lepinette08,Lepinette10}. For general stochastic volatility frameworks, this approach
 is no longer valid because $k$-th moments of the asset prices $S$ which is not in general a martingale may be infinite for $k>1$, see \cite{And-Pit,Lions-Mus}. We come over this difficulty by applying a truncation technique. In particular, for any $L>0$, we consider the stopping time 
\begin{equation}\label{sec:Discr.1}
\tau^{*}=
\tau^{*}_\zs{L}=\inf
\left\{t\ge 0: {\bf 1}_\zs{\{t\ge t^*\}}{\eta_t}^{-1}+\bar{\sigma}_\zs{t}>L
\right\}\wedge 1\,,
\end{equation}
where $\eta_t=\eta(S_t)$ and $\bar{\sigma}_t
=\max\{ \sigma(y_t), \vert \sigma'(y_t)\vert\}.$ Note that jumps may not be fully controlled for stopped process $S_\zs{t\wedge \tau^*}$ as in \cite{Nguyen}. Therefore, in the presence of jumps we consider its version defined by 
\begin{equation}
S_\zs{t}^{*}=S_{0}\exp \left\{ \int_{0}^{t} b_s\, \d s+\int_{0}^{t}{\sigma}_s^* \d W^{(1)}_\zs{s}
+\sum_{j=1}^{N_\zs{t}}\ln
\left( 1+\xi _{j}\right) \right\},
\end{equation}%
where 
$\check{b}^{*}_\zs{t}=b_\zs{t}-\sigma^{*2}_\zs{t}/2-\theta\E\,\xi_\zs{1}$
and
${\sigma}^*_t={\sigma}\left( y_\zs{t}\right){\bf 1}_\zs{\{ {\sigma}\left( y_\zs{t}\right)\le L\}}$. Here the dependency on $L$ is dropped for simplicity. Then, it is clear that $S_t^*=S_t$ on the set $\{\tau^*=1\}$. We easily observe that under Condition $(\C_\zs{2})$, 
\begin{equation}\label{sec:Est.4-00-1}
 \lim_\zs{L\to\infty}\limsup _\zs{t^*\to 1}\P ( \tau^{*}<1)=0\,.
\end{equation}
For simplicity, in the sequel we use the notation $\breve{S}_u=(S_u,y_u)$. We carry out an approximation procedure for a class of continuously differentiable functions $A:\bbr_\zs{+}\times\bbr_\zs{+}\times \bbr\to \bbr$ satisfying the following technical condition, which is more general than the one proposed in \cite{Nguyen}.

\vspace{1mm}
{\em
\noindent $(\H)$ Let $A$ be a $\bbr_+\times \bbr_+\times\bbr\to \bbr$ continuously differentiable function having absolutely integrable derivative $A'$ with respect to the first argument and for any $x>0, y\in \bbr$,
$$
\lim_\zs{n\to\infty} {n}^{\beta}\,
\left( \int_\zs{0}^{l_\zs{\ast}}|A(\lambda,x,y)|\d \lambda
+\int_\zs{l^{*}}^{+\infty } |A(\lambda,x,y)|\d \lambda\right)=0\,.
$$
Furthermore, there exist $\gamma>0$ and a positive continuous function
$U$ such that 
\begin{equation}\label{sec:Est.4-00-02}
\vert A(\lambda,x,y)\vert\le (\lambda^{-\gamma}+1) U(x,y)  \wh{\phi}(\lambda,x),
\end{equation}
where $\wh{\phi}$ defined in \eqref{sec:Bat.4-1} and 
\begin{equation}\label{sec:Est.4-00-03}
\sup_{0\le t\le 1}\E\, {U}^{4}(\breve{S}^{*}_\zs{t})<\infty.
\end{equation}
}

\begin{remark}
 In approximation of hedging error, the function $U(x,y)$ takes the form $\sqrt{x}[c_1\sigma(y)+c_2 \sigma'(y)]^m$ (up to a multiple constant) for some constants $c_1,c_2$ and $m\ge 0$. Therefore, for any $L>0$, Condition \eqref{sec:Est.4-00-03} is fulfilled as long as $\sup_\zs{0\le t\le 1}\E\,(S^{*}_\zs{t})^{2}<\infty$ but this is guaranteed by the condition of finite second moment of jump sizes $(\C_1)$. See Appendix \ref{App.Ineq}.
\end{remark}
\noindent For some positive constant $L$, we introduce the function
\begin{equation}\label{sec:Est.4-00-030}
g^*(x)={g}_L^*(x)=\vert x\vert {\bf 1}_\zs{\{\vert x\vert > L^{-1}\}}+L^{-1} {\bf 1}_\zs{\{\vert x\vert \le L^{-1}\}}.
\end{equation}
Putting $\eta_t^*=g^*(\eta_t)$, one observes that on the set $\{\tau^*=1\}$, 
\begin{equation}\label{sec:Est.4-00-0301}
\eta_\zs{t}^*= L^{-1}\quad \mbox{and}\quad\wh{\phi}(\lambda,S_t)=\phi(\lambda,\eta_t^*)={\phi}(\lambda,L^{-1}):=\phi_L(\lambda), \quad\mbox{for all}\quad t^*\le u<1.
\end{equation}

\subsection{Approximation for stochastic integrals}
For the completeness of representation we recall here the asymptotic result established in \cite{Nguyen}, which serves the central role in the proof of the main results.

\begin{proposition}\label{Pr.Discr.1} Let $A(\lambda,x,y)$ be a function such that $A$ and its first partial derivatives $\partial_xA$, $\partial_y A$ satisfy $(\H)$. Then, for $i=1,2$,
 \begin{equation}\label{sec:Est.4}
  \int_\zs{0}^1\wh{\sigma}_\zs{t}^2
\left(\int_\zs{t}^1\, 
A(\lambda_\zs{t},\breve{S}_\zs{u})\d W_\zs{u}^{(i)}\right) \d t
=\varrho^{-1}\sum_\zs{j=m_\zs{1}}^{m_2}\ov{A}_\zs{j-1}
\,{Z}_\zs{i,j}
\Delta \lambda_\zs{j}+o(n^{-\beta}),
 \end{equation}
where $\ov{A}_\zs{j}=\ov{A}(\lambda_\zs{j},\breve{S}_\zs{t_\zs{j}})$
and $\ov{A}(\lambda,x,y)=\int_{\lambda}^\infty A(z,x,y)\d z$.
\end{proposition}
\proof We follow the argument used in Proposition 7.1 in \cite{Nguyen}. Although we are working under the technical condition $(\H)$ which is slightly different from that in \cite{Nguyen}, the arguments are similar. For the reader's convenience let us present the proof in detail since the approximation technique will be repeatedly used in our analysis. First, making use of the stochastic Fubini theorem one gets
$$
\wh{I}_\zs{n}=
\int_\zs{0}^1\wh{\sigma}_\zs{t}^2
\left(\int_\zs{t}^1\, 
A(\lambda_\zs{t},\breve{S}_\zs{u})\d W_\zs{u}^{(i)}\right) \d t
=
\int_\zs{0}^1
\left(
\int^{u}_\zs{0}\,
\wh{\sigma}_\zs{t}^2\,A(\lambda_\zs{t},\breve{S}_\zs{u})\d t
\right)
\,\d W_\zs{u}^{(i)}\,.
$$
Changing the variables $v=\lambda_\zs{t}$ for the inner integral, we obtain
$$\int^{u}_\zs{0}\,
\wh{\sigma}_\zs{t}^2\,A(\lambda_\zs{t},\breve{S}_\zs{u})\d t=
 \int^{\lambda_\zs{0}}_\zs{\lambda_\zs{u}}\,A(v,\breve{S}_\zs{u})\d v
=\ov{A}(\lambda_\zs{u},\breve{S}_\zs{u})-\ov{A}(\lambda_\zs{0},\breve{S}_\zs{u}).$$ In other words,
$
\wh{I}_\zs{n}=\wh{I}_\zs{1,n}
-
\wh{I}_\zs{2,n}$, where 
$\wh{I}_\zs{1,n}=\int_\zs{0}^1\,\breve{A}_\zs{u}\,\d W_\zs{u}^{(i)}$, 
$\breve{A}_\zs{u}=\ov{A}(\lambda_\zs{u},\breve{S}_\zs{u})$
and $\wh{I}_\zs{2,n}
=\int_\zs{0}^1\,\ov{A}(\lambda_\zs{0},\breve{S}_\zs{u})\,\d W_\zs{u}^{(i)}$. 
Moreover, we have
\begin{equation}\label{sec:Est.4-00}
\wh{I}_\zs{1,n}=\int_\zs{0}^{t^\ast{*}} \breve{A}_\zs{u} \d W_\zs{u}^{(i)}+
\int_\zs{t^*}^{t_\zs{*}} \breve{A}_\zs{u} \d W_\zs{u}^{(i)}
+\int_\zs{t_\zs{*}}^{1} \breve{A}_\zs{u} \d W_\zs{u}^{(i)}
:=R_\zs{1,n}+ R_\zs{2,n}+R_\zs{3,n}\,.
\end{equation}
Let use  first show that $\wh{I}_\zs{2,n}=o(n^{-\beta})$. For any $\e>0$, one observes that
$$
\P( n^{\beta}\vert \wh{I}_\zs{2,n}\vert >\e)
\le \P( n^{\beta}\vert \wh{I}_\zs{2,n}\vert >\e, \, 
\tau^{*}=1) + \P(\tau^{*}<1).
$$
In view of \eqref{sec:Est.4-00-1}, one needs to show that the first probability in the right side converges to 0.  Indeed, by $(\H)$ one has
$$
|\ov{A}(\lambda_\zs{0},x,y)|\le C U(x,y)\int^{\infty}_\zs{\lambda_\zs{0}}e^{-\lambda/8}\d \lambda \le C {U}(x,y)e^{-\lambda_\zs{0}/8}.
$$
Putting $\breve{A}^{*}_\zs{u}=\breve{A}_\zs{u}{\bf 1}_\zs{\{\tau^{*}=1\}}$
and $\wh{I}^{*}_\zs{2,n}=\int_\zs{0}^1\,\breve{A}^{*}_\zs{u}\,\d W_\zs{u}^{(i)}$ and making use of the notation $\breve{S}^{*}=(S^*,y)$ one has
\begin{align*}
\P( n^{\beta}\vert \wh{I}_\zs{2,n}\vert >\e, \, \tau^{*}_\zs{L}=1)&=
\P( n^{\beta}\vert \wh{I}^{*}_\zs{2,n}\vert >\e)
\le \e^{-2}n^{2\beta}\E\,(\wh{I}^{*}_\zs{2,n})^2\\[2mm]
&\le  C \e^{-2}n^{2\beta}e^{-\lambda_\zs{0}/8} 
\sup_{0\le t\le 1}\,
\E \,{U}^2(\breve{S}^{*}_\zs{t}),
\end{align*}
which converges to zero by Condition $(\H)$. Hence,
$\wh{I}_\zs{2,n}=o(n^{-\beta})$ as $n\to\infty$. 
Next, let us show that $R_\zs{2,n}$ is the main part of $\wh{I}_\zs{1,n}$. For this aim, taking into account that 
$l^{*}\le \lambda_\zs{u}\le \lambda_\zs{0}$ for $0\le u\le t^{*}$,
we get $R_\zs{1,n}=o(n^{-\beta})$. 

Next, let us show the same property for the last 
term $R_\zs{3,n}$ in \eqref{sec:Est.4-00}. To this end, note again that
\begin{equation}\label{sec:Est.4-1}
 \P \left(n^{\beta}\vert R_\zs{3,n}\vert >\varepsilon\right)
\le 
\P \left(n^{\beta}\vert R_\zs{3,n}\vert >\varepsilon, \, 
\tau^*=1 \right)+\P\left(\tau^*<1\right).
\end{equation}
On the set $\{\tau^*=1\}$ one has the estimate
$
\vert\breve{A}_\zs{u}\vert \le U(\breve{S}^{*}_\zs{u})\int_\zs{\lambda_\zs{u}}^{\infty} (1+z^{-\gamma})\wh{\phi}(z,S_\zs{u}^{*})  \d z
={U}(\breve{S}^{*}_\zs{u}) \breve{f}^{*}_u,
$
 where $\breve{f}^{*}_u=\int_\zs{\lambda_\zs{u}}^{\infty} (1+z^{-\gamma}) \phi_L(z) \d z$. 
Again one obtains by the Chebychev inequality
$$
\P \left(n^{\beta}\vert R_\zs{3,n}\vert >\varepsilon, \tau^*=1 \right)=\P \left(n^{\beta}\vert \wh{R}_\zs{3,n}\vert >\varepsilon\right)\le {n^{2\beta}}{\varepsilon ^{-2}}\int_\zs{t_\zs{*}}^{1} \E (\wh{A}^{*}_\zs{u})^2 \d u,
$$
which is bounded by ${n^{2\beta}}{\varepsilon ^{-2}} C(U)
\int_\zs{t_\zs{*}}^{1} (\breve{f}^{*}_u)^2\d u$ with $C(U)=\sup_\zs{0\le u\le 1}\E\,{{U}^2(\breve{S}_\zs{u^{-}}^{*})}<\infty$.
Taking into account that 
$$\int_\zs{t_\zs{*}}^{1} (\breve{f}^{*}_u)^2\d u
={\lambda_\zs{0}^{-4\beta}}\int_\zs{0}^{l_\zs{*}}\left(\int_\zs{\lambda}^{\infty} (1+z^{-\gamma})\phi_L(z) \d z \right)^2 \d \lambda\le C{\lambda_\zs{0}^{-4\beta}} l_\zs{*},
$$
we conclude that $\lim_\zs{n\to\infty} \P \left(n^{\beta}\vert R_\zs{3,n}\vert >\varepsilon, \,\tau^*=1\right)=0$ and hence $R_\zs{3,n}=o(n^{-\beta})$ in view of \eqref{sec:Est.4-00-1}.

It remains to discretize the integral term $R_\zs{2,n}$ using the sequence $(Z_{i,j})$. 
The key steps for this aim are the following. First, we represent
$
R_\zs{2,n}=\int_{t^*}^{t_*} \breve{A}_\zs{u}\d W_\zs{u}^{(i)}=
\sum_{j=m_\zs{1}}^{m_\zs{2}}\int_{t_{j-1}}^{t_j} 
\breve{A}_\zs{u}\d W_\zs{u}^{(i)}.
$
and replace the It\^o integral in the last sum with 
$\ov{A}_\zs{j-1} Z_{i,j}\sqrt{\Delta t_j}$.
 Next, Lemma \ref{Le.Tool.1} allows to substitute $\sqrt{\Delta t_j}={\varrho}^{-1}\Delta \lambda_j$ into the last sum to obtain the martingale ${\mathcal M}_{m_\zs{2}}$ defined by
$$
{\mathcal M}_k={\varrho}^{-1}\sum_{j=m_\zs{1}}^{k}\ov{A}_{j-1} Z_{i,j}\Delta \lambda_j,\, m_\zs{1}\le k\le m_\zs{2}.
$$
We need to show that
$
\P-\lim_{n\to\infty} n^{\beta}\vert R_\zs{2,n} -{\mathcal M}_{m_\zs{2}}\vert=0
$
or equivalently, $
\sum_{j=m_\zs{1}}^{m_\zs{2}} B_\zs{j,n}=o(n^{-\beta}),
$
where $B_\zs{j,n}=\int_{t_{j-1}}^{t_j}
\wt{A}_\zs{u,j}\d W_\zs{u}^{(i)}$ and $\wt{A}_\zs{u,j}=\bar{A}(\lambda_\zs{u},\breve{S}_\zs{u})-\bar{A}(\lambda_\zs{j-1},\breve{S}_\zs{t_\zs{j-1}}).$
We show this without using the It\^o's formula. For this aim, let $b>0$ and introduce the set
$$
\Omega_\zs{b}=\left\{\sup_\zs{t^{*}\le u\le 1} \sup_\zs{z\in \bbr} \left(\vert A(z,\breve{S_\zs{u}})\vert + \left\vert {\partial}_\zs{x}  \bar{A}(z,\breve{S_\zs{u}})\right\vert +\left\vert {\partial}_\zs{y}  \bar{A}(z,\breve{S_\zs{u}})\right\vert\right)\le b\right\}.
$$
Then, for any $\varepsilon>0$, $\P\left( n^{\beta}\vert \sum_{j=m_\zs{1}}^{m_\zs{2}} B_\zs{j,n}\vert > \varepsilon\right)$ is bounded by
$$
\P(\Omega_\zs{b}^c)+\P(\tau^{*}<1)+\P\left( n^{\beta}\vert \sum_{j=m_\zs{1}}^{m_\zs{2}} B_\zs{j,n}\vert > \varepsilon, \,\Omega_\zs{b},\, \tau^{*}=1\right).
$$
Note that  $\lim_\zs{b\to\infty}\, \ov{\lim}_\zs{n\to\infty} \P(\Omega_\zs{b}^c)=0$ by Lemma \ref{Le.Tool.3-2}. In view of \eqref{sec:Est.4-00-1}, one needs to prove that the latter probability converges to zero. To this end, put 
$\wh{A}_\zs{u,j}=\wt{A}_\zs{u,j}{\bf 1}_\zs{\{\vert \wt{A}_\zs{u,j}\vert\le b\delta_\zs{u,j}\}}$ and $\wh{B}_\zs{j,n}=\int_{t_{j-1}}^{t_j}
\wh{A}_\zs{u,j}\d W_\zs{u}^{(i)},$
where $\delta_\zs{u,j}=\vert \lambda_u-\lambda_\zs{j-1}\vert+\vert{S}_\zs{u^{-}}^{*}-S_\zs{t_\zs{j-1}^{-}}^{*}\vert +\vert{y}_\zs{u^{-}}-y_\zs{t_\zs{j-1}^{-}}\vert$.
Then, the above probability is equal to
$
\P\left( n^{\beta}\vert \sum_{j=m_\zs{1}}^ {m_\zs{2}}\wh{B}_\zs{j,n}\vert>\varepsilon\right)$, which is smaller than $\varepsilon^{-2}n^{2\beta}\sum_{j=m_\zs{1}}^{m_\zs{2}}\E \, \wh{B}_\zs{j,n}^2
$ by the Chebychev inequality.
Clearly, $\E \,\wh{B}_\zs{j,n}^2$ is bounded by
\begin{align*}
2 b^2\int_{t_{j-1}}^{t_j}((\lambda_\zs{u}-\lambda_\zs{j-1})^2+
\E({S}_\zs{u}^{*}-S_\zs{t_\zs{j-1}}^{*})^2+\E({y}_\zs{u}-y_\zs{t_\zs{j-1}})^2)\d u
\le C \left((\Delta \lambda_j)^3+ (\Delta t_j) ^2\right).
\end{align*}
Consequently, $n^{2\beta}\sum_{j=m_\zs{1}}^{m_\zs{2}}\E \,\wh{B}_\zs{j,n}^2\le Cn^{2\beta}\sum_{j=m_\zs{1}}^{m_\zs{2}} (\Delta \lambda_j)^3+ (\Delta t_j)^2.$ Taking into account Lemma \ref{Le.Tool.1}  we conclude that the latter sum converges to 0 hence, the proof is completed.\endproof


\begin{lemma}\label{Le. Use.3} Let $\iota(t)=\sup\{t_i: t_i\le t\}$ and ${A}(\lambda,x,y)$ is a function satisfying Condition $(\H)$. Then,
\begin{itemize}
\item [{\bf(i)}.]
$\int_0^1\left(\int_\zs{\iota(t)}^t \wh{\sigma}_u^2A(\lambda_u,\breve{S}_\zs{u})\d u\right)\d W^{(i)}_t =o(n^{-\beta}),\quad i=1,2,$\\
\item [{\bf(ii)}.]$\int_0^1\left(\int_\zs{\iota(t)}^t A(\lambda_u,\breve{S}_\zs{u})\d W_u^{(i)}\right)\d W_t^{(j)} =o(n^{-\beta}),\quad i,j\in\{1,2\}$.
\end{itemize} 
\end{lemma}
\proof By assumption, 
$\vert A(\lambda,x,y)\vert \le  U(x,y) \wh{\phi}(\lambda,x)(1+{\lambda}^{-\gamma})$
for some constant $\gamma$ and positive function $U(x,y)$ verifying \eqref{sec:Est.4-00-03}. Denote by ${\r}_\zs{n}$ the double stochastic integral in ({\bf i}). Put $\wt{A}_t=\int_\zs{\iota(t)}^t \wh{\sigma}_u^2A(\lambda_u,\breve{S}_\zs{u})\d u$, we represent ${\r}_\zs{n}$ as
$$
{\r}_\zs{n}=\int_0^{t^*}\d \wt{A}_tW^{(i)}_t
+\int_{t^*}^1 \wt{A}_t\d W^{(i)}_t:={\r}_\zs{1,n}+{\r}_\zs{2,n}.
$$
We will prove that ${\r}_\zs{i,n}= o(n^{-\beta}),\, i=1,2$. To this end, let $L>0$ and consider $\tau^*=\tau_L^*$ defined as in \eqref{sec:Discr.1}. For $i=1,2$, by ${\r}_\zs{i,n}^*$ we mean the "corrected" version of ${\r}_\zs{i,n}$, i.e. $S_u, y_u$ are replaced by $S_u^*$ and $y_u^*$ respectively in $A$. Now, for any $\varepsilon>0$, 
\begin{equation}\label{sec:Est.4-020}
\P \left(n^{\beta}\vert {\r}_{n}\vert >\varepsilon\right)\le \P \left(n^{\beta}\vert {\r}_{n}\vert >\varepsilon, \tau^*=1\right)+\P(\tau^*<1)\,.
\end{equation}
Taking into account $\lambda_t\ge l^*\to\infty$ for $t\in[0,t^*]$ and using Chebychev's inequality, one bounds the first probability in the right side by
$$n^{2\beta} \varepsilon^{-2}\E {\r}_{1,n}^{*2}= n^{2\beta} \varepsilon^{-2} \int_0^{t^*}\E \wt{A}_t^{*2} \d t
\le C n^{2\beta} \varepsilon^{-2}\E U^2(\breve{S}^*_\zs{t}) \int_0^{t^*} \b_t^2 \d t\,,
$$
where $\b_t=\int_\zs{\iota(t)}^t  \wh{\sigma}_u^2 (1+\lambda_u^{-\gamma}) e^{-\lambda_u/8}\d u$. Recall from \eqref{Sec.Mol.6} that
\begin{equation}\label{sec:Est.4-02}
\wh{\sigma}^2_u=\varrho\sqrt{n}(1-u)^{\frac{1-\mu}{2\mu}}=\varrho\sqrt{n}(\lambda_0/\lambda_u)^{\wh{\mu}},\quad \mbox{with}\quad \wh{\mu}=(\mu-1)/(1+\mu).
\end{equation}
Then, splitting the integral as the sum of integrals on the intervals $[t_\zs{i-1},t_\zs{i}]$ and changing variable one gets 
\begin{align*}
n^{2\beta}\int_0^{t^*} \b_t^2 \d t&\le C n^{2\beta}n^{-2}\int_0^{t^*} \wh{\sigma}_u^4 (1+\lambda_u^{-\gamma})^2 e^{-\frac{\lambda_u}{4}}\d u\\[2mm]
&\le C n^{2\beta-3/2}\lambda_0^{\wh{\mu}}\int_0^{t^*} \wh{\sigma}_u^2 \lambda_u^{-\wh{\mu}}(1+\lambda_u^{-\gamma})^2 e^{-\frac{\lambda_u}{4}}\d u,
\end{align*}
which is smaller up to some constant than $n^{2\beta-3/2}\lambda_0^{\wh{\mu}}\int_{l^*}^\infty \lambda^{-\wh{\mu}}(1+\lambda^{-\gamma})^2 e^{-\frac{\lambda}{4}}\d \lambda.$ This implies that the convergence to zero of the first probability in the right side of \eqref{sec:Est.4-020}. In view of \eqref{sec:Est.4-00-1}, one obtains ${\r}_\zs{1,n}= o(n^{-\beta})$. Let us prove the same property for ${\r}_\zs{2,n}$. In fact, the singularity at $t=1$ requires a more delicate treatment. We make use of the stopping time $\tau^*$ again. Put 
$
\wh{A}^{*}_u=A(\lambda_u,\breve{S}_u^{*}){\bf 1}_\zs{\{\vert A_u\vert\le U(\breve{S}^{*}_\zs{u}) \hat{f}^*_u\}},$  $\wh{f}^*_u=(1+\lambda_\zs{u}^{-\gamma}) \phi_L(\lambda)
$
and $\wh{\r}_{2,n}=\int_{t^*}^1\left(\int_\zs{\iota(t)}^t \wh{\sigma}_u^2 \wh{A}_u^*\d u\right)\d W^{(i)}_t $. Then, by the Chebychev inequality one gets $\P \left(n^{\beta}\vert {\r}_{2,n}\vert >\varepsilon, \tau^*=1\right)
=\P \left(n^{\beta}\vert \wh{\r}_{2,n} \vert >\varepsilon\right)$. The latter probability is bounded by
\begin{align*}
{n^{2\beta}}{\varepsilon ^{-2}}\sup_\zs{0\le u\le 1}\E\,{U^2(\breve{S}_\zs{u}^{*})} \int_{t^*}^1\left(\int_\zs{\iota(t)}^t \wh{\sigma}_u^2\wh{f}^{*}_u \d u\right)^2\d t\le C{n^{2\beta}}{\varepsilon ^{-2}}\int_{t^*}^1 (\iota(t)-t)\int_\zs{\iota(t)}^t \wh{\sigma}_u^4\wh{f}^{*2}_u \d u\d t:=\a_n.
\end{align*}
On the other hand, for some constant  $C_\zs{\varepsilon,\varrho}$ independent of $n$,
$$
\a_n\le C n^{2\beta-\frac{3}{2}}{\varepsilon ^{-2}} \varrho \lambda_0^{\wh{\mu}}\int_\zs{t^*}^1 \lambda_u^{-\wh{\mu}}\wh{\sigma}_u^2\wh{f}^{*2}_u \d u \le C_\zs{\varepsilon,\varrho}n^{\frac{-2}{1+\mu}}\int^{l^*}_0 \lambda^{-\wh{\mu}}(1+{\lambda_\zs{u}}^{-\gamma})^2 \phi_L^2(\lambda)\d \lambda,
$$
which converges to 0 as $n\to\infty$. Hence, by taking into account \eqref{sec:Est.4-00-1} one concludes that $\P \left(n^{\beta}\vert {\r}_{2,n}\vert >\varepsilon\right)$ converges to 0. The second equality can be proved by the same way.\endproof

\begin{lemma}\label{Le.Use.5} 
Suppose that $A=A(\lambda,x,y)$ satisfies Condition $(\H)$. Then, the following asymptotic properties hold in probability:
\begin{itemize}
\item[{\bf (i)}.] $\int_0^1\left(\int_0^t A(\lambda_t,\breve{S}_\zs{s}) \d W_s^{(i)}\right)\d W_t^{(j)}=O(n^{-2\beta}),\quad i,j\in\{1,2\}.$
\item[{\bf(ii)}.] $\int_0^1 A(\lambda_t,\breve{S}_\zs{t})\d t=O(n^{-2\beta}).$
\item [{\bf(iii)}.] $\int_0^1\left(\int_t^ 1A(\lambda_t,\breve{S}_\zs{s})\d s\right)\d t =O(n^{-4\beta}).$
\end{itemize}
\end{lemma}
\proof
The procedure used in the proof of Lemma \ref{Le. Use.3} can be applied straightforwardly to obtain the first equality. Indeed, we can check directly that
$$\int_\zs{[0,t^{*}]\cup [t_\zs{*},1]}\left(\int_0^ tA(\lambda_t,\breve{S}_\zs{s})\d W^{(i)}_s\right)\d W_t^{(j)}=o(n^{-2\beta}).
$$
Now, consider again the set $\{\tau^*=1\}$ one can prove that
$n^{2\beta}\int_\zs{t^{*}}^{t_\zs{*}}\left(\int_0^ tA (\lambda_t,\breve{S}^*_\zs{s})\d W^{(i)}_s\right)\d W_t^{(j)}$ is bounded in probability using again the truncation technique hence, {\bf(i)} is verified. 
Next, let us prove {\bf(iii)}. By making use of the change of variable $\lambda_t=\lambda_0(1-t)^{1/(4\beta)}$, the double integral is written as
$$\wh{\epsilon}_n:=16 \lambda_0^{-8\beta} \beta^2\int_0^{\lambda_0}\lambda^{4\beta-1}\left(\int_\lambda^{\lambda_0} z^{4\beta-1}A(z,\breve{S}_\zs{v(z/ \lambda_0)})\d z\right)\d \lambda,
\quad v(z)=1-z^{4\beta}.
$$
By hypothesis, $A(\lambda,x,y)$ is bounded by ${U}(x,y)(1+\lambda^{-\gamma})\wh{\phi}(\lambda,x)$ for some constant $\gamma$ and some positive function $U$ satisfying \eqref{sec:Est.4-00-03}. Hence, $\lambda_0^{8\beta}\vert \wh{\epsilon}_n\vert$ is bounded (up to a multiple constant) by the double integral
$$
 \int_0^{\lambda_0}\lambda^{4\beta-1}\left(\int_\lambda^{\lambda_0} z^{4\beta-1}(1+z^{-\gamma}) U(\breve{S}_\zs{v(z/ \lambda_0)^{-}})\wh{\phi}(z, {S}_\zs{v({z}/{\lambda_0})})\d z\right)\d \lambda.
$$
Let $\omega$ be outside the set $\{ {S}_\zs{1}=K \}$, which has zero probability by Lemma \ref{Le. Use.1}. It is clear that the integrand of the above integral is dominated by a continuous function depending on $\omega$, which exponentially decreases to 0 at 0 and infinity hence, it is integrable on $[0,\infty)$. Therefore, the double integral converges to $$\int_0^{\infty}\lambda^{4\beta-1}U(\breve{S}_\zs{1})\left(\int_\lambda^{\infty} z^{4\beta-1}(1+z^{-\gamma}) \wh{\phi}(z, {S}_\zs{1})\d z\right)\d \lambda$$ by the dominated convergence theorem.  Thus, $n^{4\beta}\wh{\epsilon}_n$ is bounded in probability. The equality {\bf(ii)} is proved by the same way.  \endproof


\subsection{Eliminating jumps}
In this subsection, we establish asymptotic results which will serve in eliminating jump effects in our approximation. 
\begin{lemma}\label{Le. Use.4} Suppose that 
$$
\vert A(\lambda, x,y,z)\vert \le \varpi(z)\psi(\lambda) U(x,y), \quad \mbox{for all}\quad x>0, z\in \bbr, \lambda>0,
$$
where $\Pi(\varpi+\varpi^{2})<\infty$, $U$ is a continuous function satisfying $\sup_{0\le t\le 1}\E\, U^2(\breve{S}_\zs{t}^*)< \infty$ for any $L>0 $ in the definition of $\tau^*$ in \eqref{sec:Discr.1}. Suppose furthermore that
 \begin{equation}
 n^r\int_{l^*}^\infty \lambda^{4\beta-1}(\psi^2(\lambda)+\psi(\lambda))\d \lambda\to 0, \quad\mbox{for any}\; r>0.
\label{sec:Asyp.1-1-01}
 \end{equation}
Then, for any $r>0$,
\begin{equation}\label{sec:Est.04-0-1}
\int_0^{1}\int_\zs{\bbr_\zs{*}}A(\lambda_t,{S}_\zs{t^{-}}, y_\zs{t^{-}},z){J}(\d t\,,\,\d z) =o(n^{-r}).
\end{equation}
\end{lemma}

\proof For notation simplicity, one abbreviates $B(t,z):=\vert A(\lambda_t,{S}_\zs{t^{-}}, y_\zs{t^{-}},z)\vert$.
Let us decompose the integral in \eqref{sec:Est.04-0-1} as 
\begin{equation}\label{sec.Proof.2-2-00}
\int_{0}^{t^*}\int_\zs{\bbr_\zs{*}}B(t,z){J}(\d t\,,\, \d z)+\int^{1}_{t^*}\int_\zs{\bbr_\zs{*}}B(t,z){J}(\d t\,,\, \d z).
\end{equation}
Using the representation
\eqref{Sec.Mol.2}, we conclude that for any $\delta>0$ and $r>0$, the probability
$
\P\left(n^r\left \vert \int^{1}_{t^*}\int_\zs{\bbr_\zs{*}}B(t,z){J}(\d t\,,\, \d z)\right\vert>\delta\right)$ is smaller than
$\P(N_{1}-N_{t^*}\ge 1)= 1-e^{-\theta (1-t^*)},
$
which converges to $0$, when $t^*$ goes to $1$.  Hence, it suffices to prove the same property for the first integral in \eqref{sec.Proof.2-2-00}. Indeed, this term can be represented as
$$
\int_{0}^{t^*}\int_\zs{\bbr_\zs{*}}B(t,z)\wt{J}(\d t\,,\, \d z)+\int_{0}^{t^*}\int_\zs{\bbr_\zs{*}}B(t,z) \d t\,\Pi( \d z)\,.
$$
We recall that $\wt{J}(\d t\,,\, \d z)=J(\d t\,,\, \d z)-\d t\Pi( \d z)$.
We prove now  that the last term is almost surely exponentially negligible, i.e.
for any $r>0$
\begin{equation}
\label{sec.Proof.2-1-1}
\lim_\zs{n\to\infty}\,
n^r \int_{0}^{t^*}\int_\zs{\bbr_\zs{*}}B(t,z) \d t\,\Pi( \d z)=0
\quad\mbox{a.s}
\,.
\end{equation}
Indeed, by assumption and the change of variable defined in \eqref{sec:Bat.3}, it is estimated
by
$$
4\,\Pi(\varpi)\,\beta\,
\max_\zs{0\le t\le 1} U(\breve{S}_\zs{t})
\,
 n^r\lambda_0^{-4\beta}\int_{l^*}^{\infty} \lambda^{4\beta-1}\psi(\lambda)  \d \lambda
\,,
$$
 which a.s. converges to zero due to \eqref{sec:Asyp.1-1-01} and the continuity of $U$, where $t(\lambda)=1-(\lambda/\lambda_0)^{4\beta}.$
Hence, it remains to prove that for any $r>0$,
$\int_{0}^{t^*}\int_\zs{\bbr_\zs{*}} B(t,z)\wt{J}(\d t\,,\, \d z)=o(n^{-r})$ in probability as $n\to\infty$.  
To this end, note that 
$
 B(t,z)\le U(\breve{S}_\zs{t^{-}}^*) \psi(\lambda_t) \varpi(z):=\wt{B}^*(t,z)
$
on the set $\{\tau^*=1\}$, i.e. $B(t,z)=B(t,z){\bf 1}_{\{ \vert B(t,z)\vert \le \wt{B}^*(t,z)\}}:=\check{B}(t,z)$ on this set.
So, for any $\delta>0$ and $L>0$ and using the  Chebyshev inequality
we obtain that
\begin{align*}
\P\left(n^r\left\vert \int_{0}^{t^*}\int_\zs{\bbr_\zs{*}}B(t,z)\wt{J}(\d t\,,\, \d z)\right\vert>\delta\right)
&\le
\P(\tau^*<1)+
\P\left(n^r \left \vert \int_{0}^{t^*}\int_\zs{\bbr_\zs{*}}\check{B}(t,z)\wt{J}(\d t\,,\, \d z)\right\vert>\delta\right)
\\[2mm]
&
\le 
\P(\tau^*<1)+
\frac{n^{2r} \E\, \left( \int_{0}^{t^*}\int_\zs{\bbr_\zs{*}}\check{B}(t,z)\wt{J}(\d t\,,\, \d z)\right)^{2}}{\delta^{2}}
\,.
\end{align*}
Using here the inequality \eqref{Novikov++} for $p=2$ 
and taking into account that
$$
\E \int_{0}^{t^*}\int_\zs{\bbr_\zs{*}}\wt{B}^{*2}(t,z)\Pi(\d z) \d t
\le
\Pi(\varpi^2)
\sup_\zs{0\le t\le 1}\E\, U^2( \breve{S}_\zs{t}^*) \int_{0}^{t^*}\psi^2(\lambda_t) \d t 
\,
 \Pi(\varpi^2)
 \le 
 C 
 \int_{l^*}^{\infty} \lambda^{4\beta-1}\psi(\lambda)  \d \lambda
 \,,
$$
 the limit equation
\eqref{sec:Est.04-0-1}
 follows from \eqref{sec:Asyp.1-1-01}. \endproof

\subsection{Limit theorems for approximations}\label{CTL}
We first recall the following result in \cite{Hall}, which is useful for studying asymptotic distribution of discrete martingales. 

\begin{theorem} \label{Th.sec:Bat.0}[Theorem 3.2 and Corollary 3.1, p.58 in \cite{Hall}]
 Let ${\mathcal M}_\zs{n}=\sum_\zs{i=1}^{n}X_\zs{i}$ be a zero-mean, square
integrable martingale and $\varsigma $ be an a.s. finite random variable.
Assume that the following convergences are satisfied in probability:
$$
\sum_\zs{i=1}^{n}\E\left( X_\zs{i}^{2}\boldsymbol{1}_\zs{\left\{ \left\vert
X_\zs{i}\right\vert >\delta \right\} }|\mathcal{F}_\zs{i-1}\right) \longrightarrow
0\quad\mbox{for any}\quad \delta >0\quad \mbox{and}\quad \sum_\zs{i=1}^{n}\E\left( X_\zs{i}^{2}|\mathcal{F}_\zs{i-1}\right)
\longrightarrow \varsigma^{2}.
$$
Then, the sequence $({\mathcal M}_\zs{n})$ converges in law to $X$ whose
characteristic function is $\E\exp ( -\frac{1}{2}\varsigma
^{2}t^{2}),$ i.e. $X$ has a Gaussian mixture distribution.
\end{theorem}

Below we will establish some special versions of Theorem \ref{Th.sec:Bat.0}. In particular, our aim is to study the asymptotic distribution of discrete martingales resulting from approximation \eqref{sec:Est.4} in Proposition \ref{Pr.Discr.1}.

Let $A_{i}=A_{i}(\lambda,x,y), \, i\in I:=\{1,2,3,4\}$ be functions having property $(\H)$ and 
consider discrete martingales $({\mathcal M}_\zs{k})_\zs{m_\zs{1}\le k\le m_\zs{2}}$ and $(\ov{{\mathcal M}}_k)_\zs{m_\zs{1}\le k\le m_\zs{2}}$ defined by
\begin{equation}\label{sec:Bat.4}
{\mathcal M}_\zs{k}=\varrho^{-1}\sum_\zs{j=m_\zs{1}}^k\sum_\zs{i\in I\backslash\{4\}} {A}_\zs{i,j-1}
\,{Z}_\zs{i,j}
\Delta \lambda_\zs{j}\quad \mbox{and}\quad\ov{{\mathcal M}}_k=\varrho^{-1}\sum_\zs{j=m_\zs{1}}^{k}\sum_\zs{i\in I\backslash\{3\}}{A}_\zs{i,j-1}\,{Z}_\zs{i,j}
\Delta \lambda_\zs{j},
\end{equation}
where ${A}_\zs{i,j}={A}_\zs{i}(\lambda_\zs{j},\breve{S}_\zs{t_\zs{j}})$
and $Z_{i,j}$ are defined as in \eqref{sec:Tool.12-1} and \eqref{sec.Tool.12}. To describe the limiting distributions let us introduce
\begin{align}\label{sec:Bat.4-00-1-2}
&\L=
A^{2}_\zs{1}+
2A_\zs{1}A_\zs{3}
( 2\Phi({p}) -1) 
+A^{2}_\zs{3}\,\Lambda({p}) 
+A^{2}_\zs{2},\quad
&\ov{\L}=
A^{2}_\zs{1}+A^{2}_\zs{2}
+(1-2/\pi)A^{2}_\zs{4},
\end{align}\label{sec:Bat.5}
where ${p}$ is defined in \eqref{sec:Bat.3-1}. Define now
\begin{equation}\label{sec:Bat.4-00-1}
\varsigma^{2}=\breve{\mu} \varrho ^{\frac{2}{\mu +1}
}\int_\zs{0}^{+\infty }\lambda^{\wh{\mu}}\L(\lambda,\breve{S}_\zs{1}) \d\lambda \quad\mbox{and}\quad \quad \ov{\varsigma}^{2}=\breve{\mu}\, \varrho ^{\frac{2}{\mu +1}
}\int_\zs{0}^{+\infty }\lambda^{\wh{\mu}}\,\ov{\L}(\lambda,\breve{S}_\zs{1}) \d\lambda
\end{equation}
with
\begin{equation}\label{sec:Bat.4-00-1-1}
\breve{\mu}=\frac{1}{2}(\mu +1)\wt{\mu}^{\frac{2}{\mu +1}} \quad\mbox{and}\quad \wh{\mu}=(\mu -1)/(\mu +1).
\end{equation}

\begin{proposition}\label{Pr.Limit.1}Assume that $A_{i}=A_{i}(\lambda,x,y), \, i=1,2,3$ and their first partial derivatives $\partial_\lambda A_i$, $\partial_xA_i$, $\partial_yA_i$ are functions satisfying Condition $(\H)$. Then, for any fixed $\varrho>0$ 
the sequence  
$
(n^{\beta }{\mathcal M}_\zs{m_\zs{2}})
$
 weakly converges to a 
mixed Gaussian variable with mean zero and variance $\varsigma^{2}$ defined
as in \eqref{sec:Bat.4-00-1}.
The same property still holds if some (or all) of the functions $A_i$ are replaced by $\int_{\lambda}^\infty A_{i}(z,x,y)\d z$.
\end{proposition}

\proof Note that the square integrability property is not guaranteed for the random variables 
$(\upsilon_\zs{j})$. To overcome this issue let us recall the stopping time $\tau^*=\tau^*_L$ defined in \eqref{sec:Discr.1} and put
$
\wt{A}_i(\lambda,x,y)=A_i(\lambda,x,y)\wh{\phi}^{-1}(\lambda,x){\phi}_L(\lambda),
$
where $\phi_L(\lambda)$ defined in \eqref{sec:Est.4-00-030}. Let 
 $ \upsilon_\zs{j}^{*}=\sum^{3}_\zs{i=1}\wt{A}_\zs{i}(\lambda_\zs{j},\breve{S}^{*}_\zs{t_j})\,{Z}_\zs{i,j}
\Delta \lambda_\zs{j}$ and $
{\mathcal M}_\zs{k}^{*}=\sum_\zs{j=m_\zs{1}}^k\upsilon_\zs{j}^{*}.
$

\noindent{\bf Step 1}: We will show throughout Theorem \ref{Th.sec:Bat.0} that for any $L>0$
the martingale $n^{\beta}{\mathcal M}_\zs{m_2}^{*}$ weakly converges to
a mixed Gaussian variable with mean zero and variance 
$\varsigma^{*2}(L)$ defined as
\begin{equation}
\varsigma^{*2}(L)=\breve{\mu} \varrho ^{\frac{2}{\mu +1}
}\int_\zs{0}^{+\infty }\lambda^{\wh{\mu}}\wt{\L}(\lambda,\breve{S}_\zs{1}) \d\lambda
\,,
\label{sec:Bat.5-1-1}
\end{equation}
where $\wt{\L}$ is obtained by replacing all $A_i$ in the formula of $\L$ in \eqref{sec:Bat.4-00-1-2} by the corresponding modified functions $\wt{A}_i, i=1,2,3$
. To this end,
 setting $
\a^{*}_\zs{j}=\E\,(\upsilon^{*2}_\zs{j}\,\boldsymbol{1}
_\zs{\left\{ \left\vert \upsilon^{*}_\zs{j}\right\vert >\delta\right\}} \vert {\cal F}_\zs{j-1}),
$ we first show that $\P\left(n^{2\beta} \vert\sum_\zs{j=m_\zs{1}}^{m_\zs{2}}
\a^{*}_\zs{j}\vert>\varepsilon\right)$ converges to 0. 
By hypothesis,
\begin{equation}\label{sec:Bat.5-1}
\max_{i=1,2,3}\left\vert \wt{A}_i(\lambda_\zs{u}, \breve{S}_\zs{u}^{*})\right\vert\le U(\breve{S}_\zs{u}^{*})(1+\lambda_\zs{u}^{-\gamma})\phi_L(\lambda)\le 
U(\breve{S}_\zs{u}^{*})(1+\lambda_\zs{u}^{-\gamma})
\end{equation} 
for some $\gamma>0$ and positive function $U(\breve{S})$ satisfying \eqref{sec:Est.4-00-03}. 
 We observe that
$$
\P\left(n^{2\beta} \vert\sum_\zs{j=m_\zs{1}}^{m_\zs{2}}
 \a^{*}_\zs{j}\vert>\varepsilon\right)=\P\left(n^{2\beta} \vert\sum_\zs{j=m_\zs{1}}^{m_\zs{2}}
 {\a}^{*}_\zs{j}\vert>\varepsilon\right)\le \varepsilon^{-1}n^{2\beta} \sum_\zs{j=m_\zs{1}}^{m_\zs{2}}
\E\, {\a}^{*}_\zs{j}
$$
by Markov's inequality. Using the Chebychev inequality and then again the Markov inequality one gets
\begin{align*}
\E\, \a^{*}_\zs{j}=\E\,\left({\upsilon}_\zs{j}^{*2}\boldsymbol{1}
_\zs{\left\{ \left\vert \upsilon_\zs{j}^{*}\right\vert >\delta\right\}}\right)&\le \sqrt{\E\,\upsilon_\zs{j}^{*4}}\sqrt{\P(\vert \upsilon_\zs{j}^{*}\vert >\delta)} \le \delta^{-2}\E\,\upsilon_\zs{j}^{*4} \\
&\le 9\delta^{-2} (1+\lambda_\zs{u}^{-\gamma})^4(\Delta \lambda_\zs{j})^4 \E \,U^4(\breve{S}_\zs{u}^{*})\sum_{i=1}^3
\,{Z}^4_\zs{i,j}.
\end{align*}
Taking into account that all of $Z_\zs{i,j}$ have bounded moments and using \eqref{sec:Bat.5-1}
we obtain 
$$
\varepsilon^{-1}\,n^{2\beta}\, \sum_\zs{j=m_\zs{1}}^{m_\zs{2}}
\E\, \a^{*}_\zs{j}\le 9C\varepsilon^{-1}\delta^{-2} n^{2\beta}\,
 \sum_\zs{j=m_\zs{1}}^{m_\zs{2}}
(1+\lambda_\zs{u}^{-\gamma})^4  (\Delta \lambda_\zs{j})^4,
$$
which converges to 0 by Lemma \ref{Le.Tool.1}.

Let us verify the limit of the sum of conditional variances $\E(\upsilon_\zs{j}^{*2}|\mathcal{F}_\zs{j-1})$. Setting $\upsilon_\zs{i,j}^{*}=\wt{A}^{*}_\zs{i,j-1}
\,{Z}_\zs{i,j}\,\Delta \lambda_\zs{j}$, one obtains
$
\E\left(\upsilon_\zs{1,j}^{*}\upsilon_\zs{3,j}^{*}\vert\mathcal{F}_\zs{j-1}\right) =
\E\left( \upsilon_\zs{2,j}^{*}\upsilon_\zs{3,j}^{*}|\mathcal{F}_\zs{j-1}\right)
=0
$
since ${Z}_\zs{1,j}$ and ${Z}_\zs{2,j}$ are independent.
It follows that
$$
\E( \upsilon_\zs{j}^{*2}|\mathcal{F}_\zs{j-1})=\E( \upsilon_\zs{1,j}^{*2}|\mathcal{F}_\zs{j-1})
+\E( \upsilon_\zs{2,j}^{*2}|\mathcal{F}_\zs{j-1})
+\E( \upsilon_\zs{3,j}^{*2}|\mathcal{F}_\zs{j-1})
+2\E( \upsilon_\zs{1,j}^{*}\upsilon_\zs{2,j}^{*}|\mathcal{F}
_\zs{j-1}).
$$
Observe that for $Z\sim N( 0,1) $ and some constant $a$, $
\E( Z\left\vert Z+a\right\vert) =2\Phi (a)
-1$ and $
\E\,\left( Z+a\right)^{2} -
\left(\E\vert Z+a\right\vert)^{2}=\Lambda(a),
$
where $\Phi$ is the standard normal distribution function and $\Lambda$ is defined in \eqref{sec:Bat.3-0}. On the other hand, $\Delta {
\lambda}_\zs{j}=n^{-2\beta}(
1+o( 1) ) \breve{\mu}\,\varrho ^{\frac{2}{\mu +1}}\lambda_\zs{j-1}^{\wh{\mu }}$ by Lemma \ref{Le.Tool.1}. So, 
$$
n^{2\beta }\E( \upsilon_\zs{j}^{*2}|\mathcal{F}_\zs{j-1})=(
1+o( 1) ) \breve{\mu}\,\varrho ^{\frac{2}{\mu +1}}\,
\lambda_\zs{j-1}^{\wh{\mu }}\,\wt{\L}(\lambda_\zs{j-1},\breve{S}_\zs{t_\zs{j-1}^{-}}^{*}) \Delta {
\lambda}_\zs{j}.
$$
Therefore, by Lemma \ref{Le:Bat.5}, the sum
$n^{2\beta}\,\sum_\zs{j=m_\zs{1}}^{m_\zs{2}}
\E( \upsilon_\zs{j}^{*2}|\mathcal{F}_\zs{j-1})$
converges in probability to $\varsigma ^{*2}(L)$ defined in \eqref{sec:Bat.5-1-1}.
Thus, $n^{\beta}{\mathcal M}_\zs{m_2}^{*}$ weakly converges to ${\cal N}(0,\varsigma ^{*2}(L))$ throughout Theorem~\ref{Th.sec:Bat.0}. 

\noindent{\bf Step 2}: Let us show that 
$
\sup_\zs{\epsilon>0}\lim_\zs{L\to\infty}\limsup_\zs{n\to\infty}\P\left(\vert n^{\beta}{\mathcal M}_\zs{m_2}^{*}-n^{\beta}{\mathcal M}_\zs{m_2}\vert>\epsilon\right)=0.
$
To this end, recall that $\wh{\phi}(\lambda,S_t)={\phi}_L(\lambda)$ and hence, $\wt{A}_i=A_i$ for $i=1,2,4$ on the set $\{\tau^*=1\}$. Then, the conclusion directly follows from  
$$
\P\left(n^{\beta}\vert {\mathcal M}_\zs{m_2}^{*}-{\mathcal M}_\zs{m_2}\vert>\epsilon\right)\le \P\left(n^{\beta}\vert {\mathcal M}_\zs{m_2}^{*}-{\mathcal M}_\zs{m_2}\vert>\epsilon, \tau^*=1\right)+\P(\tau^*<1)
$$
and \eqref{sec:Est.4-00-1}.
Moreover, taking into account that $\varsigma ^{*2}(L)$ converges a.s. to $\varsigma^{2}$ as $L\to\infty$, we conclude that $n^{\beta} {\mathcal M}_\zs{m_2}$ converges in law to ${\cal N}(0,\varsigma^2)$, which completes the proof.\endproof

\vspace{3mm}

Let us consider martingales of the following form, resulting from the approximation for L\'epinette's strategy,
$\ov{{\mathcal M}}_k=\sum_\zs{j=m_\zs{1}}^{k}\left({A}_\zs{1,j-1}
\,{Z}_\zs{1,j}+{A}_\zs{2,j-1}\,{Z}_\zs{2,j}
+{A}_\zs{4,j-1}
\,{Z}_\zs{4,j}\right)
\Delta \lambda_\zs{j}. 
$
Their limiting variance is defined throughout the function
\begin{align}
\ov{\L}( \lambda, x,y)=
A^{2}_\zs{1}(\lambda,x,y)+A^{2}_\zs{2}(\lambda, x,y)
+(1-2/\pi)A^{2}_\zs{4}(\lambda, x,y).
\end{align}
The following result is similar to Proposition \ref{Pr.Limit.1}.
\begin{proposition}\label{Pr.sec:Bat.3} Suppose that $A_{i}=A_{i}(\lambda,x,y), \, i=1,2,4$ and their first partial derivatives satisfying Condition $(\H)$. Then, for any fixed $\varrho>0$ 
the sequence  
$
(n^{\beta }\,\ov{{\mathcal M}}_\zs{m_\zs{2}})$
weakly converges to a 
mixed Gaussian variable with mean zero and variance $\ov{\varsigma}^{2}$ given by \eqref{sec:Bat.4-00-1}.
The same property still holds if some (or all) of the functions $A_i$ are replaced by $\int_{\lambda}^\infty A_{i}^{0}(z,x,y)\d z$.
\end{proposition}

\proof The conclusion follows directly from the proof of Proposition \ref{Pr.Limit.1} and the observation that 
$\E\, {Z}_\zs{4,j}^2=\E(\vert {Z}_\zs{1,j}\vert-\sqrt{2/\pi})^2=1-{2}/{\pi},$ and 
$ \E \,({Z}_\zs{i,j}{Z}_\zs{4,j})=0,
$ for $i=1,2$ and $m_\zs{1}\le j\le m_\zs{2}$.\endproof


\vspace{2mm}
The remaining part of the section is devoted to prove main results following the scheme of \cite{Nguyen}. Our first step is to establish the asymptotic representation at rate $n^{\beta}$ for each term contributing in the hedging error. The approximation procedure also provides the residual parts as discrete martingales for which Propositions \ref{Pr.Limit.1} and \ref{Pr.sec:Bat.3} will be applied to achieve the limiting distribution in the last step.

\subsection{Approximation for $I_\zs{1,n}$}
The following approximation is obtained in \cite{Nguyen}.
\begin{proposition}\label{Pr.Proof.1}
Let $\breve{H}=\int_\lambda^\infty(z^{-1/2}/2-z^{-3/2}\ln(x/K))\wt{\varphi}(z,x) \d z$ and define
 $$
{\cal U}_\zs{1,k}=\varrho^{-1}\sum_\zs{j=m_1}^k\sigma ( y_\zs{t
_\zs{{j-1}}}) S_\zs{t_\zs{j-1}}\,
\breve{H}_\zs{j-1}\,
 Z_\zs{1,j}\,\Delta 
{\lambda}_\zs{j}, \quad m_1\le j\le m_2.
$$
Then, under $(\C_1)$ and $(\C_2)$, 
$
\P-\lim_\zs{n\longrightarrow \infty }n^{\beta}
\left\vert I_\zs{1,n}-2\min (S_\zs{1},K)
-
{\cal U}_\zs{1,m_2}
\right\vert
=0.
$
\end{proposition}
\proof
By \eqref{Sec.Mol.14-1}, one represents $I_\zs{1,n}$ as 
$$
I_\zs{1,n}=\int_\zs{0}^{1}\,
\wh{\sigma}_\zs{t}^{2} S_\zs{t}^{2}\wh{C}_\zs{xx}( t,S_\zs{t}) \d t-\int_\zs{0}^{1}\,
\sigma ^{2}(y_\zs{t})\, S_\zs{t}^{2}\wh{C}_\zs{xx}( t,S_\zs{t}) \d t.
$$
The last term is $n^{\beta}$ negligible by $({\bf ii})$ of Lemma \ref{Le.Use.5}. To study the first integral let us introduce the function $A(\lambda,x)=x^{2}\wh{C}_\zs{xx}( t,x)$ and split it as
$$
\int_\zs{0}^{1}\,
\wh{\sigma}_\zs{t}^{2} S_\zs{t}^{2}\wh{C}_\zs{xx}( t,S_\zs{t})
 \d t=
\int_\zs{0}^{1}\,
\wh{\sigma}_\zs{t}^{2} S_\zs{1}^{2}\wh{C}_\zs{xx}( t,S_\zs{1}) \d t+\int_\zs{0}^{1}\,
\wh{\sigma}_\zs{t}^{2} \left(A(\lambda_t, S_\zs{t})-A(\lambda_t, S_\zs{1})\right)\d t.
$$
The first integral $\int_\zs{0}^{1}\,
\wh{\sigma}_\zs{t}^{2} S_\zs{1}^{2}\wh{C}_\zs{xx}( t,S_\zs{1}) \d t $ almost surely converges to $2\min(S_\zs{1},K)$ faster than $n^r$ for any $r>0$, see \cite{Nguyen}. Let us study the last term which describe jumps of $A$. Using the It\^o Lemma for $A(\lambda_t, S_\zs{t})-A(\lambda_t, S_\zs{1})$, we rewrite it as 
\begin{align}\label{sec.Proof.1}
\epsilon_\zs{1,n}+\epsilon_\zs{2,n}+
\int_\zs{0}^{1}\int_t^1\,
\wh{\sigma}_\zs{t}^{2} \partial_xA(\lambda_t, S_\zs{u}) \sigma(y_u)S_\zs{u}\d W^{(1)}_\zs{u}\d t,
\end{align}
where 
$$
\epsilon_\zs{1,n}:=
\int_\zs{0}^{1}\int_t^1\,
\wh{\sigma}_\zs{t}^{2} A_1(\lambda_t, S_\zs{u},y_\zs{u})\d u\,\d t
\quad\mbox{and}\quad
\epsilon_\zs{2,n}:=\int_\zs{0}^{1}\wh{\sigma}_\zs{t}^{2}\int_t^1\int_\zs{\bbr_\zs{*}} \bar{A}(\lambda_t, S_\zs{u^{-}},z) J(\d u\,,\, \d z) \d t
$$
with 
$$
A_1(\lambda,x,y)=\partial_\zs{t}A(\lambda, x)+\partial_\zs{xx}A(\lambda, x) \sigma^2(y)x^2, \quad 
\bar{A}(\lambda, x,z)=A(\lambda,x(1+z))-A(\lambda,x).
$$
Then, the approximation procedure of Proposition \ref{Pr.Discr.1} is used to get a discrete martingale approximation ${\cal U}_\zs{1,m_2}$ for the It\^o's integral of \eqref{sec.Proof.1}.

Now, let us show that $\epsilon_\zs{i,n}=o(n^{-\beta}), i=1,2$. In fact, $\epsilon_\zs{1,n}=o(n^{-\beta})$ by $({\bf iii})$ of Lemma \ref{Le.Use.5}. The jump term $\epsilon_\zs{2,n}$ 
can be represented as 
$
\epsilon_\zs{2,n}=\int_\zs{0}^{1}\int_\zs{\bbr_\zs{*}}\left(\int_0^u\wh{\sigma}_\zs{t}^{2}\bar{A}(\lambda_t, S_\zs{u^{-}},z) \d t\right)J(\d u\,,\, \d z)
$
by Fubini's theorem \cite{App}. Changing variable $v=\int_u^1\wh{\sigma}_\zs{t}^{2} \d t$ as in \eqref{sec:Bat.3}, one gets
$$
\int_0^u\wh{\sigma}_\zs{t}^{2}\bar{A}(\lambda_t, \cdot,z) \d t= \int_{\lambda_u}^{\lambda_0}\bar{A}(v, \cdot,z) \d v:=D(\lambda_u, \cdot,z)
$$
and hence, $\epsilon_\zs{2,n}=\int_\zs{0}^{1}\int_\zs{\bbr_\zs{*}} D(\lambda_u, S_\zs{u^{-}},z) J(\d u\,,\, \d z)$.
 On the other hand, for any $\Upsilon>0$
$$
 D(\lambda_u, \Upsilon,z)=\int_\zs{\lambda_u}^{\lambda_0}\bar{A}(v, \Upsilon,z)
 \d v
 =\int_\zs{\Upsilon}^{\Upsilon(1+z)} \int_{\lambda_u}^{\lambda_0}\partial_x A(v,\x) \d v \d \x.
$$
Direct computation shows that $\partial_x A(v,\x)=2 \x \wh{C}_\zs{xx}(v,\x)+\x^2 \wh{C}_\zs{xxx}(v,\x)$ and 
$$
\wh{C}_\zs{xx}(v,\x)=\frac{1}{\x\sqrt{v}}\wt{\varphi}(v,\x),\quad \wh{C}_\zs{xxx}(v,\x)=-\frac{1}{\x^2v}\wt{\varphi}(v,\x)\left(\frac{3}{2}\sqrt{v}+\frac{\ln(\x/K)}{\sqrt{v}}\right),
$$
where
$$
\wt{\varphi}(v,\x)=\frac{1}{\sqrt{\x}}\phi_0(v) e^{-\frac{\ln^2(\x/K)}{2 v}}\quad \mbox{with}\quad 
\phi_0(v)=\sqrt{\frac{K}{2\pi}}e^{-v/8}\,.
$$
Using the fact that $z^k e^{-z^2/2}$ is uniformly bounded for all $k$, one has
$$
\vert  \partial_x A(v,\x)\vert \le C\frac{1}{\sqrt{\x}} (1+{v}^{-1})\phi_0(v),
$$
for some positive constant $C$.
This estimate implies that for any $\Upsilon>0$
\begin{equation}\label{sec.Proof.1-1-1}
\vert D(\lambda_u, \Upsilon,z)\vert\le C\left \vert \int_\zs{\Upsilon}^{\Upsilon(1+z)}\frac{1}{\sqrt{\x}}  \d \x \right\vert \,
 \int_\zs{\lambda_u}^{\lambda_0} (1+{v}^{-1})\phi_0(v) \d v \le C \vert z\vert \wt{\phi}_0(\lambda_u)  \sqrt{\Upsilon}\,,
\end{equation}
where 
$\wt{\phi}_0(\lambda)=\int_\lambda^\infty (1+{v}^{-1})\phi_0(v) \d v$.
Clearly, Condition \eqref{sec:Asyp.1-1-01} in Lemma \ref{Le. Use.4} holds, hence,
$\epsilon_\zs{2,n}=o(n^{-r})$ for any $r>0$.\endproof

\subsection{Approximation for $I_\zs{2,n}$}
\begin{proposition}\label{Pr.Proof.2}
Under $(\C_1)$ and $(\C_2)$, $n^{\beta}I_\zs{2,n}$ converges to 0 in probability as $n\to\infty$.
\end{proposition}
\proof
We represent $I_\zs{2,n}$ as 
\begin{equation}\label{sec.Proof.2}
\int_\zs{0}^{1} \sigma(y_t){S}_\zs{t} A(t)\d W_t^{(1)}+ \int_\zs{0}^{1} \int_\zs{\bbr_\zs{*}} z {S}_\zs{t^{-}} A(t-)\wt{J}(\d t\,,\, \d z):=\b_\zs{1,n}+\b_\zs{2,n},
\end{equation}
where
${A}(t)=\wh{C}_x (\iota(t),{S}_\zs{\iota(t)}) - \wh{C}_x(t,{S}_\zs{t})$. We first claim that the It\^o's integral of \eqref{sec.Proof.2} can be omitted by Lemma \ref{Le. Use.3}. To see this, it suffices to apply the It\^o's formula, one represents the difference $ {A}_\zs{t}$ as
\begin{align*}
 \int_\zs{\iota(t)}^t \left(\wh{C}_\zs{xt}(u,{S}_\zs{u})+\sigma^2(y_u){S}_\zs{u}^2\wh{C}_\zs{xxx}(u,{S}_\zs{u})\right)\d u +
 \int_\zs{\iota(t)}^t \wh{C}_\zs{xx}(u,{S}_\zs{u})\sigma(y_u){S}_\zs{u}\d W_\zs{u}^{(1)}\\
+ \int_\zs{\iota(t)}^t \int_\zs{\bbr_\zs{*}} (\wh{C}_\zs{x}(u,{S}_\zs{u^{-}}(1+z))-\wh{C}_\zs{x}(u,{S}_\zs{u^{-}})) J(\d u\,,\, \d z).
\end{align*}
In view of \eqref{Sec:nocost.1}, 
\begin{equation}\label{sec.Proof.2-0}
\wh{C}_\zs{xt}(u,x)= -\frac{1}{2}\wh{\sigma}_u^2\left(2x \wh{C}_\zs{xx}(u,x)+ x^2 \wh{C}_\zs{xxx}(u,x)\right):=\wh{\sigma}_u^2 \wt{A}(u,x).
\end{equation}
Therefore, $\b_\zs{1,n}$ equals the following sum
\begin{align}
&\int_\zs{0}^{1} \int_\zs{\iota(t)}^t \wh{\sigma}_u^2 \sigma(y_t){S}_\zs{t} \wt{A}(u,{S}_\zs{u})\d u \d W_t^{(1)}+ 
\int_\zs{0}^{1}\int_\zs{\iota(t)}^t \sigma(y_t){S}_\zs{t} \sigma^2(y_u){S}_\zs{u}^2\wh{C}_\zs{xxx}(u,{S}_\zs{u})\d u \d W_t^{(1)} \notag\\
&+\int_\zs{0}^{1} \int_\zs{\iota(t)}^t \int_\zs{\bbr_\zs{*}} \sigma(y_t){S}_\zs{t} (\wh{C}_\zs{x}(u,{S}_\zs{u^{-}}(1+z))-\wh{C}_\zs{x}(u,{S}_\zs{u^{-}})) J(\d u\,,\, \d z) \d W_t^{(1)}.\label{sec.Proof.2-2-2}
\end{align}
The first two integrals converge to 0 more rapidly than $n^{-\beta}$ by Lemma \ref{Le. Use.3}. Let us study the jump term in \eqref{sec.Proof.2-2-2}, which will be denoted by $\a_\zs{n}$. Clearly, by Fubini's theorem $\a_\zs{n}$ equals 
\begin{equation}
\sum_\zs{1\le i \le n}\int_\zs{t_\zs{i-1}}^{t_i} \int_\zs{\bbr_\zs{*}} \Psi(u,{S}_\zs{u^{-}},z) \left(\int_\zs{u}^{t_i} \sigma(y_t){S}_\zs{t}\d W_t^{(1)} \right)J(\d u\,,\, \d z),\label{sec.Proof.2-2-3}
\end{equation}
where $\Psi(u,x,z):=\wh{C}_\zs{x}(u,x(1+z))-\wh{C}_\zs{x}(u,x)$. We prove that $\a_\zs{n}=o(n^{-r})$ for any $r>0$ following the demonstration of Lemma \ref{Le. Use.4} with some modification. In particular, we decompose the sum in \eqref{sec.Proof.2-2-3} into two parts: $\a_\zs{1,n},$ the first concerns the index $i$ with $m_2\le i\le n$ and the second one $\a_\zs{2,n},$ which is the sum over the rest of index $i, 1\le i\le m_2$. Clearly, $\P(n^r\vert \a_\zs{1,n}\vert <\delta)\le \P(N_1-N_\zs{t^*}\le 1)= 1- e^{-\theta(1-t^*)},$ which converges to 0. 

To study $\a_\zs{2,n}$, we run again the argument used to to obtain the estimate \eqref{sec.Proof.1-1-1}. In particular, 
$\vert \Psi(u,x,z)\vert$ is bounded by
$$
\phi_1(\lambda_u) \left \vert \int_x^{x(1+z)}\frac{\d a}{a^{3/2}} \right\vert 
\le 2
\phi_1(\lambda_u) \sqrt{x}\,
\varpi_\zs{0}(z)
\,,
$$
where 
\begin{equation}\label{sec.Proof.2-2-4-1}
 \phi_1(\lambda)=\sqrt{K/2\pi}  \lambda_t^{-1/2} e^{-\lambda/8}
 \quad \mbox{and}\quad
 \varpi_\zs{0}(z)=
 \frac{\vert z\vert}{\sqrt{1+z}}
 \,.
\end{equation}
Denote by $\a_\zs{2,n}^c$ the compensator of $\a_\zs{2,n}$. Then, it is clear that 
\begin{align}
\vert \a_\zs{2,n}^c\vert &\le \sum_\zs{1\le i \le m_2}\int_\zs{t_\zs{i-1}}^{t_i} \int_\zs{\bbr_\zs{*}} \vert \Psi(u,{S}_\zs{u},z)\vert \left\vert \int_\zs{u}^{t_i} 
\sigma(y_t) S_\zs{t}\d W_t^{(1)} \right\vert \Pi(\d z)\d u \notag
\\
&\le 
\sum_\zs{1\le i \le m_2}\int_\zs{t_\zs{i-1}}^{t_i} 
\phi_1(\lambda_u) \sqrt{{S}_\zs{u}} \left\vert \int_\zs{u}^{t_i} \sigma(y_t) 
S_\zs{t}\d W_t^{(1)} \right\vert \d u \, 
\Pi(\varpi_\zs{0})\,.
\label{sec.Proof.2-2-4}
\end{align}
Note that in view of Condition $(\C_1)$ the integral $\Pi(\varpi_\zs{0})<\infty$.
It is important to highlight that $ X_u :=\sqrt{{S}_\zs{u}} \left\vert\int_\zs{u}^{t_i} \sigma(y_t){S}_\zs{t}\d W_t^{(1)}\right\vert$ may not be squared integrable. To overcome this issue, consider the stopping time $\tau^*$ defined in \eqref{sec:Discr.1} for some $L>0$. On the set $\{\tau^*=1\}$, one has for $u\in[t_\zs{i-1}, t_i]$,
$$
\E (X^{*}_\zs{u})^{2}=\E  \left(\sqrt{{S}_\zs{u}^*} \int_\zs{u}^{t_i} \sigma(y_t^*){S}_\zs{t}^*\d W_t^{(1)}\right)^2\le 
\E S_\zs{u}^* \int_\zs{u}^{t_i} {S}^{*2}_\zs{t}\d t \le C L^2 n^{-1}
$$
and hence, $\E X_u^{*}\le \sqrt{\E (X_u^{*})^{2}}\le C L n^{-1/2}$ by Cauchy-Shwart's inequality.
Therefore, 
\begin{align}
\P( n^r\vert \a_\zs{2,n}^c\vert > \delta, \tau^*=1)\notag&\le n^r \delta^{-1} \sum_\zs{1\le i \le m_2}\int_\zs{t_\zs{i-1}}^{t_i} 
\phi_1(\lambda_u)\E X_u^{*} \d u  \,
\Pi(\varpi_\zs{0})\notag\\
&
\le n^r \delta^{-1}C L n^{-1/2}\sum_\zs{1\le i \le m_2}\int_\zs{t_\zs{i-1}}^{t_i} 
\phi_1(\lambda_u)\d u  \, 
\Pi(\varpi_\zs{0}) \notag \\
& \le 
n^r \delta^{-1}C L n^{-1/2}\int_\zs{0}^{t^*} 
\phi_1(\lambda_u)\d u  \,
 \Pi(\varpi_\zs{0}). \label{sec.Proof.2-2-6}
\end{align}
Taking into account that $n^{r}\int_\zs{0}^{t^*} 
\phi_1(\lambda_u)  \d u $  goes to $0$, one concludes that
 for any $r>0$,
$
\lim_\zs{n\to\infty}
\P( n^r\vert \a_\zs{2,n}^c\vert > \delta, \tau^*=1)
=0
$. 
Noting that 
$$
\P( n^r\vert \a_\zs{2,n}^c\vert > \delta)\le \P( n^r\vert \a_\zs{2,n}^c\vert > \delta, \tau^*=1)+\P( \tau^*<1)
$$
and using \eqref{sec:Est.4-00-1} one obtains $ n^r\a_\zs{2,n}^c\to0$ in probability for any $r>0$. 

Now, putting $\wt{\alpha}_\zs{2,n}={\alpha}_\zs{2,n}-{\alpha}_\zs{2,n}^c$, we need to show that $\P(n^r\vert \wt{\alpha}_\zs{2,n}\vert >\delta)\to 0$.
To this end, consider again the stopping time $\tau^*$ defined in \eqref{sec:Discr.1} for some $L>0$. On the set $\{\tau^*=1\}$, one has
$$
 \vert \Psi(u,S_\zs{u},z)\vert\le \sqrt{{S}_\zs{u}^*} \phi_1(\lambda_u) \varpi_\zs{0}(z)\,,
$$ 
where ${S}_\zs{u}^*$ is the stopped version of ${S}_\zs{u}$. Clearly,
$$
\sup_\zs{1\le i\le n} \sup_\zs{t_\zs{i-1}\le u\le t_i}\E \,{{S}_\zs{u}^*}\vert W_\zs{t_\zs{i-1}}^{(1)}-W_\zs{u}^{(1)}\vert^2\le C n^{-1} 
$$
for some positive constant $C$. It then follows by the Chebychev inequality that
$$
\P(n^r\vert \wt{\alpha}_\zs{2,n}\vert >\delta, \tau^*=1)\le n^{2r} \delta^{-2}
\E \,\wt{\alpha}_\zs{2,n}^{*2}\,,
$$
where $\wt{\alpha}_\zs{2,n}^{*2}$ is obtained by substituting ${S}_\zs{u}$ by ${S}_\zs{u}^*$ in the function $\Psi(u,{S}_\zs{u},z)$.
Now, the well-known isometry for jump integrals applying to $\wt{\alpha}_\zs{2,n}={\alpha}_\zs{2,n}-{\alpha}_\zs{2,n}^c$ implies that $\E\, \wt{\alpha}_\zs{2,n}^{*2}$ is bounded by
$$
\sum_\zs{1\le i \le m_2} \E \int_\zs{t_\zs{i-1}}^{t_i} \int_\zs{\bbr_\zs{*}} \vert \Psi(u,{S}_\zs{u}^*,z)\vert^2 \vert W_\zs{t_\zs{i-1}}^{(1)}-W_\zs{u}^{(1)}\vert^2 \Pi(\d z)\,
\d u\,,
$$
which is smaller than
$$
\int_\zs{0}^{t^*} 
\phi_1^2(\lambda_u) \E (X^{*}_u)^{2}  \d u  \, \int_\zs{\bbr_\zs{*}} \varpi_\zs{0}^2(z)
\Pi(\d z)\le C n^{-1}
\int_\zs{0}^{t^*} 
\phi_1^2(\lambda_u)  \d u  \, \int_\zs{\bbr_\zs{*}} \varpi_\zs{0}^2(z)\,
\Pi(\d z)\,.
$$
Again, $\int_\zs{\bbr_\zs{*}} \varpi_\zs{0}^2(z)\nu(\d z)<\infty$ by Condition $(\C_1)$. Therefore, for some constant depending on $L$, 
$$
\P(n^r\vert \wt{\alpha}_\zs{2,n}\vert >\delta, \tau^*=1)\le C(L) n^{2r} \delta^{-2} n^{-1}
\int_\zs{0}^{t^*} 
\phi_1^2(\lambda_u)  \d u  \times \int_\zs{\bbr_\zs{*}} \varpi_\zs{0}^2(z)\Pi(\d z)\,,
$$
which converges to 0 for any $r>0$. Letting now $L\to\infty$ and using \eqref{sec:Est.4-00-1} we obtain that $\vert \wt{\alpha}_\zs{2,n}\vert =o(n^{-r})$ for any $r>0$.
By the same way, one can show that $n^r\b_\zs{2,n}\to 0$ in probability for any $r>0$ and the proof is completed.\endproof

\subsection{Approximation for $I_\zs{3,n}$}

\begin{proposition}\label{Pr.Proof.3}
Suppose that $(\C_1)$ and $(\C_2)$ hold. Then, for any $r>0$, $n^{r}\vert I_\zs{3,n}\vert\to 0$ in probability as $n\to\infty$.
\end{proposition}
\proof By \eqref{Sec.Mol.14-1}, one has
$
\B(t,S_\zs{t^{-}},z)=\int_\zs{S_\zs{t}}^{S_\zs{t}(1+z)} \int_\zs{S_\zs{t}}^v\wh{C}_{xx}(t,u) \d u \d v.
$
Recall that 
$
\wh{C}_\zs{xx}(t,u)=u^{-1} \lambda_t^{-1/2} \wt{\varphi}(\lambda_t,u)\le u^{-3/2} \phi_1(\lambda_t),
$
where $\phi_1(\lambda)=\sqrt{K/(2\pi)}  \lambda_t^{-1/2} e^{-\lambda/8}$.
Direct calculus leads to 
$
\vert \B(t,S_\zs{t},z)\vert \le C S_\zs{t}^{1/2} \phi_1(\lambda_t) \vert z\vert. 
$
Therefore all assumptions in Lemma \ref{Le. Use.4} are fulfilled and the conclusion follows. \endproof

\subsection{Approximation for $\Gamma_\zs{n}$}

Let us study the trading volume $\Gamma_n$. It is easy to check that for $v\ge 0,\ 1-\Phi (v)\le Cv^{-1}\varphi(v)$ and 
$\int_0^{t^{*}} \wt{\varphi}(\lambda_u,S_u)\d u+\int_{t_{*}}^1 \wt{\varphi}(\lambda_u,S_u)\d u$ 
almost surely converges to 0 more rapidly than any power of $n$. 
Therefore, one can truncate the sum and keep only the part corresponding to index $m_1\le j\le m_2$. Next, one can ignore jumps terms that may appear in approximations via It\^o's formulas in the interval $[t^*, 1]$. For convenience, let us recall here the approximation result for $\Gamma_n$ obtained in \cite{Nguyen}.
\begin{proposition}\label{Pr.sec:Trc.2}
Under Conditions $(\C_1)-(\C_2)$, the total trading volume
$\Gamma_{n}$ admits the following asymptotic form
$$
\Gamma_\zs{n}=\Gamma({S}_\zs{1},{y}_\zs{1},\varrho)+({\cal U}_\zs{2,m_2}+{\cal U}_\zs{3,m_2})+o(n^{-\beta}).
$$
\end{proposition}


\subsection{Proof of Theorem \ref{Th.sec:Mol.1}}
By Propositions \ref{Pr.Proof.1}-\ref{Pr.sec:Trc.2}, the hedging error is represented as $V^n_1-h(S_1)=\min \left(
{S}_{1^{-}},K\right)-\kappa \Gamma\left( {S}_{1^{-}},y_{1},\varrho\right)+{\cal M}_\zs{m_2}$, where 
the martingale part of the
hedging error is given by ${\cal M}_\zs{k}=\frac{1}{2}{\cal U}_\zs{1,k}-\kappa({\cal U}_\zs{2,k}+{\cal U}_\zs{3,k})$ and hence, the sequence $\left(n^{\beta }{\cal M}_\zs{m_2}\right)$
converges in law to a mixed Gaussian variable by Proposition~\ref{Pr.Limit.1}
and Theorem~\ref{Th.sec:Mol.1} is proved.\endproof

\subsection{Proof of Theorem \ref{Th.sec:Mol.2}}
Suppose now that the L\'epinette strategy $\bar{\gamma} _\zs{t}^{n}$ is applied for the replication problem. Analogously one can represent the corresponding hedging error as
$
\bar{V}^{n}_\zs{1}-h(S_\zs{1})=\frac{1}{2}I_\zs{1,n}+\bar{I}_\zs{2,n}-I_\zs{3,n}-\kappa \bar{\Gamma}_n,
$
where $$\bar{I}_\zs{2,n}={I}_\zs{2,n}+\sum_{i\ge 1}\Delta S_\zs{t_{i}}\int_\zs{0}^{t_{i-1}}\wh{C}_\zs{xt}(u,S_\zs{u})\d u$$ and $
\bar{\Gamma}_n=\sum^{n}_\zs{i=1}\,S_\zs{t_\zs{i}}\vert \bar{\gamma}_\zs{t_\zs{i}}^{n}-\bar{\gamma}
_\zs{t_\zs{i-1}}^{n}\vert\,
$ is the trading volume. Recall that
${I}_\zs{2,n}$ est negligible by Proposition \ref{Pr.Proof.2}. Let us investigate the above sum. By \eqref{sec.Proof.2-0}, it can be represented as 
$$
\sum_{i\ge 1}\int^{\lambda_0}_{\lambda_{i-1}}\wt{A}(u,S_\zs{u})\d v \, \int_{t_{i-1}}^{t_{i}}\sigma(y_\zs{t}) S_\zs{t}\d W^{(1)}_\zs{t}
+\sum_{i\ge 1}\int^{\lambda_0}_{\lambda_{i-1}}\wt{A}(u,S_\zs{u})\d v \,\int_{t_{i-1}}^{t_{i}}z\d S_\zs{t^{-}} \wt{J}(\d t\,,\,\d z)
$$
using the above change of variable, where $\wt{A}$ is defined in \eqref{sec.Proof.2-0}. Now, the approximation technique of Proposition \ref{Pr.Discr.1} can be applied to replace the first sum by the martingale $\ov{\cal U}_\zs{2,m_2}$ defined by
$$
\ov{{\cal U}}_\zs{2,k}=\varrho^{-1}\sum_\zs{j=m_1}^k\sigma ( y_\zs{t
_\zs{{j-1}}}) S_\zs{t_\zs{j-1}^{-}}\,
Y_\zs{j-1}\,
 Z_\zs{1,j}\,\Delta 
{\lambda}_\zs{j},\quad m_1\le k\le m_2
$$ and $Y(\lambda,x)=\int_\lambda^\infty z^{-3/2}\ln(x/K)\wt{\varphi}(z,x)\d z$.
On the other hand, one obtains the same estimate \eqref{sec.Proof.1-1-1} for the integrand, which implies that the second sum can be omitted at order $n^r$ for any $r>0$ by Lemma \ref{Le. Use.4}.

Now, we consider the approximation representation for the trading volume $\ov{\Gamma}_\zs{n} $ following the procedure in the approximation of $\Gamma_n$. The following is established in \cite{Nguyen}. 
\begin{proposition}\label{Pr.sec:Lepi.1}
Under Conditions $(\C_1)-(\C_2)$, 
$$
\P-\lim_\zs{n\rightarrow \infty }n^{\beta}\vert\ov{\Gamma}_\zs{n}-\eta\min(S_1,K)-(\ov{\cal U}_\zs{2,m_2}+\ov{\cal U}_\zs{3,m_2})\vert=0.$$
\end{proposition}
\noindent Hence, $\ov{{\cal M}}_\zs{m_\zs{2}}={\cal U}_\zs{1,m_\zs{2}}+\ov{{\cal U}}_\zs{1,m_\zs{2}}-\kappa_\zs{*}(\ov{ \cal U}_{2,m_\zs{2}}+\ov{ \cal U}_{3,m_\zs{2}})$ is the martingale part of the
hedging error for L\'epinette's strategy, which can be represented in the form
$$
\ov{{\cal M}}_\zs{k}={\varrho }^{-1}
\sum_\zs{j=m_1}^{k}({A}_\zs{1,j-1}Z_\zs{1,j}+
{A}_\zs{4,j}Z_\zs{4,j-1}+{A}_\zs{2,j-1}Z_\zs{2,j})\Delta {\lambda}_\zs{j}
$$
for some explicit functions ${A}_\zs{i}$ satisfying the assumption of Proposition~\ref{Pr.sec:Bat.3}. Then, the convergence in law to a mixed Gaussian variable of the sequence $\left(n^{\beta }\ov{{\cal M}}_\zs{m_2}\right)$ is guaranteed by Proposition~\ref{Pr.sec:Bat.3}
and hence, Theorem~\ref{Th.sec:Mol.2} is proved.\endproof

\subsection{Proof of Theorem \ref{Th.Extension.1}}
Note first that the approximation representation for the replication error is the same as in the SVJ case. In particular, the approximations of $I_\zs{i,n},\,i=1,2,3$ are the same since the martingale sums are obtained by one-dimension It\^o's formula. The only difference is that in finding the limit of the total transaction costs one has replaced $S_\zs{t^{-}_{j-1}}$ and $y_\zs{t^{-}_{j-1}}$ by terminal values $S_\zs{1^{-}}$ and $y_\zs{1^{-}}$. Now, the two-dimension version of It\^o's formula applied for the difference provides the sums concerning the dynamics of $y_t$. By an elementary property of Poisson process one can ignore the jump part of $y_t$ in the time interval $[t^*,1]$. Hence, the martingale approximation for this difference is the same as in the SVJ case. However one needs to check the integrability of $\alpha_i(t,y_t), i=1,2$. For this aim, the condition $\sup_\zs{0\le t\le 1}\E\, y_t^2<\infty$ is needed but this is fulfilled under Condition $(\C_3)$ together with the linear growth and Lipshitz properties of these coefficients, see Appendix \ref{App.SDEJ}. \endproof

\subsection{Proof of Theorem \ref{Th.Const.1}}
The proof can be proceeded in a similar way to that for Theorem \ref{Th.sec:Mol.1} and Theorem \ref{Th.sec:Mol.2} but with a simpler argument. In fact, the difference between the use of $\wh{\sigma}^2=\sigma^2+ \varrho_0 \sqrt{n f'(t)}$ and that of the simple form $\wh{\sigma}^2=\varrho_0 \sqrt{n f'(t)}$ only comes from the approximation due to the substitution $ \wh{\lambda}=\int_t^1\wh{\sigma}_s^2 \d s$. In particular,
$$
\wh{\lambda}_t=\sigma^2(1-t)+\varrho\sqrt{n}\int_t^1 \sqrt{f'(t)}=\sigma^2(1-t)+\lambda_t,
$$ 
where $\lambda_t$ is defined by the same formula with the simple form. Then, Lemma \ref{Le.Tool.1} is still true for the sequence $(\wh{\lambda}_{j})$ constructed from the classical form \eqref{eq:const.1}.
Note that the index $m_1,m_2$ are now replaced by $\wh{m}_1$, $\wh{m}_2$ which are defined by
$$
\wh{m}_\zs{1}=n-[n \vartheta^{-1}(l^{*})],\quad 
\wh{m}_\zs{2}=n-[n \vartheta^{-1}(l_{*})],
$$
where $\vartheta (z)= \sigma^2 z^{\mu}+\lambda_0 z^{(1+\mu)/2}$, which is an increasing function for $\mu\ge 1$. Here
the notation $[x]$ stands for the integer part of a number $x$ and $
l_\zs{*}=\ln^{-3}n,\ l^{*}=\ln^{3}n
$. Similarly, we consider a subsequence $(t_j)$  of trading times 
and the corresponding sequence
$\left(\wh{\lambda}_\zs{j}\right)$
 defined as 
\begin{equation}\label{sec:Bat.3-111-1}
t_\zs{j}=1-(1-j/n) ^{\mu }
\quad\mbox{and}\quad
\wh{\lambda}_\zs{j}=\sigma^2(1-{t_j})+ \lambda_\zs{0}(
1-t_\zs{j})^\frac{1}{4\beta},\quad {\wh{m}_\zs{1}\le j\le \wh{m}_\zs{2}}.
\end{equation}
The rest of the approximation is the same as that of Theorem \ref{Th.sec:Mol.1} and Theorem  \ref{Th.sec:Mol.2}.\endproof

\section{Concluding remark}\label{Examples}
Diffusion-based stochastic volatility models well account for volatility clustering, dependence in increments and long term smiles and skews but can not capture jumps or realistic short-term implied volatility patterns. These shortcomings can be fixed by adding jumps into the model. In this paper, we contribute to the field of approximate hedging under transaction costs using Leland's algorithm by allowing for jumps. We showed that jumps in such frameworks do not affect asymptotic property of the replication error in approximate hedging with transaction costs. As a direct implication of our main result, we confirm that for constant cost rate, the Kabanov-Safarian-Pergamenshchikov results in \cite{Kab-Saf1,Per} also hold for jump-diffusion settings. It would be interesting to investigate the asymptotic properties of jump risk in small transaction costs models. 


\vspace{2mm}

\noindent{\bf Acknowledgements}: This work was partially supported by the Ministry of Education and Science of the Russian Federation (research project No.~$2.3208.2017/4.6$ and the Russian Federal Professor program  No. 1.472.2016/1.4) and by 
"The Tomsk State University competitiveness improvement programme"\ Grant 8.1.18.2018.



\section*{Appendix}\label{sec:A}
\setcounter{section}{0}
\renewcommand{\thesection}{\Alph{section}}
\section{Auxiliary Lemmas}\label{sec:Appendix.1}
\begin{lemma}\label{Le.Tool.1} There exist two positive constants $C_1,C_2$ such that
\begin{equation}\label{sec:Est.1-2}
 C_1\,{n}^{-2\beta}\varrho^{\frac{2}{\mu+1}}\,\nu_0(l_*)\le \inf_\zs{m_\zs{1}\le j\le m_\zs{2}}\vert\Delta\lambda_j\vert\le
\sup_\zs{m_\zs{1}\le j\le m_\zs{2}}\vert\Delta\lambda_j\vert\le C_2{n}^{-2\beta}\varrho^{\frac{2}{\mu+1}}\,\nu_0(l^*),
 \end{equation}
  where $\nu_0(x)=x^{(\mu-1)/(\mu+1)}$. Moreover, let $\breve{\mu}=\frac{1}{2}(\mu +1)\wt{\mu}^{\frac{2}{\mu +1}},$ then
 \begin{equation}\label{sec:Est.1-1}
  \Delta\lambda_j=\breve{\mu} {n}^{-2\beta}\varrho^{\frac{2}{\mu+1}}\,\nu_0(\lambda_{j-1})(1+o(1))\quad\mbox{and}\quad
{\Delta\lambda_j}\,(\Delta t_j)^{-1/2}=\varrho (1+o(1)).
\end{equation}
\end{lemma}
\proof It follows directly from relation \eqref{sec:Bat.3}.\endproof

\begin{lemma} \label{Le. Use.1}
For any  $K>0$ and $0<t\le 1$,
$\P(S_t=K)=0.$
\end{lemma}
\proof We prove that for $0<t\le 1$ and any real number $a$, $\P(\psi_t=a)=0,$ where
$\psi_\zs{t}=\int_{0}^{t} b_t\, \d t+\int_{0}^{t}\sigma \left( y_{s}\right) \d W^{(1)}_{s}-\frac{1}{2}\int_{0}^{t}\sigma ^{2}\left( y_{s}\right) \d s+\sum_{j=1}^{N_\zs{t}}\ln
\left( 1+\xi _{j}\right).$ Indeed, one can represent $W^{(1)}_\zs{t}= \rho B_t+ \sqrt{1-\rho^2}Z_t$, where $B_t$ is the Brownian driving $y_t$ and $Z$ is another Brownian independent of $B$. Now, conditionally on the Brownian $B$ and jump terms $\sum_{j=1}^{N_\zs{t}}\ln\left( 1+\xi_{j}\right)$, $\psi_t$ is a Gaussian variable.\endproof

\begin{lemma}\label{Le. Use. 2}For any $K>0$, $\lim_\zs{\varepsilon\to 0}\limsup_\zs{v\to 1} \P(\inf_\zs{v\le u\le 1} \vert \ln(S_u/K)\vert < \varepsilon )=0.$
\end{lemma}
\proof  For any $\varepsilon>0$, the probability in the lemma is bounded by
\begin{align*}
\P(\inf_\zs{v\le u\le 1} \vert \ln({S}_\zs{u}/K)\vert < \varepsilon , N_1-N_v=0,\psi_v^*\le\varepsilon)+\P( \psi_v^*>\varepsilon,N_1-N_v=0)+2\P(N_1-N_v\ge1),
\end{align*}
where $\psi_v^*=\sup_\zs{v\leq u\leq 1}\left\vert \ln {S_u}/{S_1}\right\vert$.
Noting $\psi_v^*\to 0$ almost everywhere in the set $\{N_1-N_v=0\}$ as $v\to1$ and $\P(N_1-N_v\ge1)=1-e^{-\theta (1-v)}\to 0$ as $v\to1$ we get
\begin{align}
 \limsup_\zs{v\to 1} \P(\inf_\zs{v\le u\le 1} \vert \ln(S_u/K)\vert<\varepsilon )&\le \limsup_\zs{v\to 1}\P(\inf_\zs{v\le u\le 1} \vert \ln({S}_\zs{u}/K)\vert < \varepsilon , N_1-N_v=0,\psi_v^*\le\varepsilon)\notag\\
&\le \P(\vert \ln(S_1/K)\vert \le 2\varepsilon).\label{eq:limsup}
\end{align}
To see \eqref{eq:limsup} we remark that for any $u\in[v,1]$, 
 \begin{align*}
 \left\vert \ln (S_\zs{1}/K) \right\vert\le \left\vert \ln (
 S_\zs{1}/S_u)\right\vert+ \left\vert \ln (S_\zs{u}/K) \right\vert
 \le\varepsilon+\left\vert \ln (
 S_\zs{u}/K) \right\vert
 \end{align*}
on the set $\{\inf_\zs{v\le u\le 1} \vert \ln(S_u/K)\vert <\varepsilon,\psi_v^*\le\varepsilon\}$. 
 Therefore, $$\left\vert \ln (S_\zs{1}/K) \right\vert\le \varepsilon+\inf_\zs{v\le u\le 1} \vert \ln(S_u/K)\vert\le 2\varepsilon.$$
Hence,
$$\P(\inf_\zs{v\le u\le 1} \vert \ln(S_u/K)\vert<\varepsilon )\le\P(\vert \ln(S_1/K)\vert \le 2\varepsilon)+\P( \psi_v^* >\varepsilon)$$
and \eqref{eq:limsup} is obtained by taking into account $\psi_v^*\to 0$ a.s. as $v\to 1$.  Letting now $\varepsilon\to0$ we get $\lim_\zs{\varepsilon\to 0}\P( \vert \ln({S}_\zs{1}/K)\vert \le 2\varepsilon)=\P({S}_\zs{1}=K)=0$ by Lemma \ref{Le. Use.1}, which proves Lemma \ref{Le. Use. 2}.\endproof

\begin{lemma}\label{Le.Tool.3-2}Suppose that $A=A(\lambda,x,y)$ and its partial derivatives 
$\partial_\lambda A,\partial_xA, \partial_yA$ satisfy Condition $(\H)$. Set 
$$
r_\zs{n}=\sup_\zs{(z,r,d)\in [l_\zs{*}, l^{*}]\times{\cal B}}
\left(\vert \partial_\zs{\lambda}{A}(z,r,d)\vert +\vert \partial_\zs{x}{A}(z,r,d)\vert+\vert\partial_\zs{y}{A}(z,r,d)\vert\right),
$$ where
${\cal B}=[S_\zs{\min},S_\zs{\max}]\times [y_\zs{\min},y_\zs{\max}]$ with
$$
S_\zs{\min}=\inf_\zs{t^*\le u\le t_*}S_\zs{u},\quad  S_\zs{\max}=\sup_\zs{t^*\le u\le t_*}S_\zs{u},\quad 
y_\zs{\min}=\inf_\zs{t^*\le u\le t_*}y_\zs{u},\quad
y_\zs{\max}=\sup_\zs{t^*\le u\le t_*}y_\zs{u}\,.
$$ Then, $\lim_\zs{b\to \infty} \ov{\lim}_\zs{n\to\infty} \P( r_n >b)=0.$
\end{lemma}
\proof See Lemma A.4 in \cite{Nguyen} with the remark that the left continuity of $S_\zs{t^{-}}$ and $y_\zs{t^{-}}$ gives the same argument. \endproof


\begin{lemma}\label{Le:Bat.5}
Suppose that $A=A(\lambda,x,y)$ and its first partial derivatives satisfy Condition $(\H)$. Set $\ov{A}(\lambda,x,y)=\int_\lambda A(z,x,y)\d z$ and $\wt{A}(\lambda,x,y)=\ov{A}^2(\lambda,x,y).$ Then, for any $\gamma>0$,
$$\P-\lim_\zs{n\to\infty}\left\vert\sum_{j=m_1}^{m_2}\lambda_\zs{j-1}^\gamma\wt{A}(\lambda_\zs{j-1}, \breve{S}_\zs{t_\zs{j-1}})
\Delta \lambda_j-\int_0^\infty \lambda^\gamma\wt{A}(\lambda, \breve{S}_\zs{1}) \d\lambda\right\vert=0,
$$ where $\breve{S}_\zs{t}=({S}_\zs{t},y_\zs{t})$. The same property holds if $\ov{A}(\lambda,x,y)=A(\lambda,x,y)$ or the product of these above functions.
\end{lemma}
\proof See Lemme A.5 in \cite{Nguyen}.\endproof

\section{Some moment estimates}\label{App.Ineq}
\begin{lemma}\label{Le.App.Ine.1}
Let $y_t$ be an It\^o's process and $S_t$ be the asset process given by 
$$
S_\zs{t}=S_{0}\exp \left\{ \int_{0}^{t} b_t\, \d t+\int_{0}^{t}\sigma \left( y_{s}\right) \d W^{(1)}_{s}-\frac{1}{2}\int_{0}^{t}\sigma ^{2}\left( y_{s}\right) \d s\right\}\prod_{j=1}^{N_\zs{t}}
\left( 1+\xi _{j}\right) ,
$$
where
$N_t$ is a homogeneous Poisson process with intensity $\theta$ independent of $(\xi_j)_\zs{j\ge 1}$ (a sequence of i.i.d. variables).  We assume that the jumps ingredient $(\xi_j)_\zs{j\ge 1}$ and $N_t$ are independent of the Brownian motion $W^{(1)}$ and of that of $y$. If $b$ and $\sigma$ are two bounded functions then, for any $m>0$,
for which $\E(1+\xi_1)^m<\infty$, we have
$$
\E S_t^m\le C(m) \exp\{ \theta t(\E(1+\xi_1)^m-1)\}, \quad \mbox{for all}\quad t\in[0,1],
$$
where $C(m)$ is some constant depending on $m$.
\end{lemma}
\proof Let us represent $S_t=\wt{b}_t {\cal E}_t(\sigma)X_t$ with $X_t=\prod_{j=1}^{N_\zs{t}}
\left( 1+\xi_{j}\right)$ and
$$
\wt{b}_t=S_{0}e^{\int_{0}^{t} b_s\, \d s}, \quad {\cal E}_t(\sigma)= \exp\left\{\int_{0}^{t}\sigma \left( y_{s}\right) \d W^{(1)}_{s}-\frac{1}{2}\int_{0}^{t}\sigma ^{2}\left( y_{s}\right) \d s\right\}.
$$
By hypothesis, the stochastic exponential ${\cal E}_t(\sigma)$ is a martingale with expectation 1, indepedent of $X_t$. Therefore,
$
\E S_t^m \le C \E {\cal E}_t^m(\sigma) \E X_t^m
$
since $\sup_\zs{0  \le t\le 1} \wt{b}_t^m\le C$. Because $\sigma$ is bounded one has
$$
\E\, {\cal E}_t^m(\sigma) =\E\, {\cal E}_t(m\sigma) e^{(m^2-m)/2 \int_0^t \sigma^2(y_s) \d s}\le C\, \E\, {\cal E}_t(m\sigma)=C.
$$
On the other hand, by the usual conditioning technique we get
$$
\E X_t^m=\E \prod_{j=1}^{N_\zs{t}}
\left( 1+\xi _{j}\right)^m= \exp\{ \theta t(\E(1+\xi_1)^m-1)\},
$$
which implies the desired conclusion. \endproof

\begin{lemma}\label{Le.App.Ine.2} Under the assumptions of Lemma \ref{Le.App.Ine.1}, for $0\le u\le v\le 1$, 
$$
\E (S_u-S_v)^{2}\le C \vert u-v\vert,
$$
for some constant $C$.
\end{lemma}
\proof
For $0\le u<v\le 1$, put $\wt{b}_\zs{v/u}:=e^{\int_{u}^{v} b_s}$, $X_\zs{u/v}:= \prod_{j=N_u+1}^{N_{v}}
\left( 1+\xi_{j}\right)$ and
$$
{\cal E}_\zs{v/u}(\sigma):=\exp\left\{\int^{v}_{u}\sigma \left( y_{s}\right) \d W^{(1)}_{s}-\frac{1}{2}\int^{v}_{u}\sigma ^{2}\left( y_{s}\right) \d s\right\}.
$$
Then, ${\cal E}_\zs{v/u}(\sigma)$ and $X_\zs{u/v}$ are independent and 
\begin{equation}\label{eq:App.Ine.1}
\sup_\zs{0\le u\le v\le 1}(\E {\cal E}_\zs{v/u}^2(\sigma)+\E e^{2\int_{u}^{v} b_s})<\infty 
\end{equation}
since $b$ ans $\sigma$ are bounded. Denote $\delta =\theta( v-u)$. It is easy to check that 
\begin{equation}
\E (X_\zs{u/v}-1)^2=e^{ \delta (\E\xi_1+\E\xi_1^2)}-2e^{\delta \E\xi_1}+1.
\label{eq:App.Ine.2}
\end{equation}
Let us first show that 
$\E (X_\zs{u/v}-1)^2\le C \delta,
$ for some constant $C$. Obviously, for any finite interval $[a,b]$, $\vert e^x-1\vert \le C x$ by Taylor's approximation. From  Condition $(\C_3)$, $\E\xi_1^2<\infty$. Now, if $\E \xi_1=0$ then $\E (X_\zs{u/v}-1)^2=e^{ \delta \E\xi_1^2}-1\le C\delta$. Similarly, in case $\E\xi_1+\E\xi_1^2=0$ one has $\E \xi_1\neq 0$ and hence $\E (X_\zs{u/v}-1)^2=e^{ \delta \xi_1}-1\le C\delta$. Lastly, if both $\E \xi_1$ and $\E\xi_1+\E\xi_1^2$ are non zero one can estimate $\E (X_\zs{u/v}-1)^2$ by
$
\vert e^{ \delta(\E \xi_1+ \E\xi_1^2 }-1\vert +2 \vert e^{ \delta \E \xi_1}-1\vert \le C\delta.
$
Using the same argument one can easily prove that
\begin{equation}
\E ({\cal E}_\zs{u/v}(\sigma)-1)^2\le C \delta \quad \mbox{and}\quad\E (\wt{b}_\zs{v/u}-1)^2\le C \delta^2.
\label{eq:App.Ine.3}
\end{equation}
Clearly, $
\E (S_u-S_v)^{2}=\E S_u^2 \E\left(\frac{S_v}{S_u}-1\right)^{2}
$
and 
\begin{equation}
\label{eq:App.Ine.4}
\left(\frac{S_u}{S_v}-1\right)^{2}\le 2 \left(\wt{b}_\zs{v/u}^2( {\cal E}_\zs{u/v}(\sigma)-1)^2+(\wt{b}_\zs{v/u}-1)^2+\wt{b}_\zs{v/u}^2{\cal E}_\zs{u/v}^2(\sigma)(X_\zs{u/v}-1)^2\right).
\end{equation}
By Lemma \ref{Le.App.Ine.1}, $\sup_\zs{0\le u\le 1}\E S_v^2<\infty$. Now, taking expectation in \eqref{eq:App.Ine.4} and using \eqref{eq:App.Ine.1}, \eqref{eq:App.Ine.2} and \eqref{eq:App.Ine.3} one obtains the conclusion.\endproof

\section{Stochastic differential equations with jumps}\label{App.SDEJ}
In this section we recall the basic result in the theory of stochastic differential equations with jumps (SDEJ) of the form
\begin{equation}
\d y_\zs{t}=\alpha_1\left( t,y_{t^{}}\right) \d t+\alpha_2\left( t,y_{t^{}}\right) \d W_\zs{t}+\d \zeta_\zs{t},
\label{eq:SDEJ.1}
\end{equation}
on the time interval $[0,T]$ with initial value $y_0$, where the process $\zeta_\zs{t}$ defined in
\eqref{jumps-1}
is independent of the Brownian motion $(W_\zs{t})_\zs{t\ge 0}$ and $\E y_0^2<\infty$. 

We first recall the Novikov inequalities \cite{Novikov1975} (also referred to as the Bichteler--Jacod inequalities, see 
\cite{BichtelerJacod1983, MarinelliRockner2014}) which provides upper bounds of the moments of the supremum of purely discontinuous local martingales:
\begin{equation}
\label{Novikov++}
\E\sup_\zs{0\le t\le n}|\Upsilon*(J-\nu)_\zs{t}|^{p}\le C_\zs{p}
 \left(
 \E\,\big (|\Upsilon|^{2}*\nu_\zs{n}\big)^{p/2}
 +
 \E\,\big (|\Upsilon|^{p}*\nu_\zs{n}\big)
\right)\,,
\end{equation}
for $p\ge 2$ and for any $n\ge 0$, where $C_\zs{p}$ is some positive constant. 

\begin{theorem}
Assume that  $\alpha_i,\,i=1,2$ are locally Lipschitz and linearly bounded functions, 
$\E y_0^2<\infty$ 
and that Condition \eqref{levy-measure_condition} holds.
 Then, there exists a unique solution $y_t$ to \eqref{eq:SDEJ.1} with initial value $y_0$ and 
\begin{equation}\label{eq:SDEJ.1-1}
 \E\,(\sup_\zs{0\le t\le T}  y_t)^2<C(T) (1+\E \,y_0^2)<\infty.
\end{equation}
Furthermore, for any $0\le s\le t\le T$, there exists a positive constant $C$ such that 
\begin{equation}\label{eq:SDEJ.2}
 \E \vert y_t-y_s\vert^2\le C \vert t-s\vert.
\end{equation}
\end{theorem}
\proof The existence and uniqueness of the solution follows by adapting the classical method used for SDEs, see for instance Theorem 2.2 in \cite{Fried}. To prove \eqref{eq:SDEJ.2}, we note that
$$
\E\vert y_t-y_s\vert^2 \le 3 \E\left(\int_s^t \alpha_1\left( u,y_{u}\right) \d u\right)^2+3\int_s^t \E \alpha_2^2\left( u,y_{u}\right) \d u + 3
 \E \left(
 \zeta_\zs{t}
 -
 \zeta_\zs{s}
  \right)^2.
$$
By the linear boundedness of $\alpha_1, \alpha_2$ and \eqref{eq:SDEJ.1-1} one gets
$$
\E\left(\int_s^t \alpha_1\left( u,y_{u}\right) \d u\right)^2\le C\vert t-s\vert \int_s^t (1+\E y_u^2) \d u\le C\vert t-s\vert ^2.
$$
Similarly, $\E \alpha_2^2\left( u,y_{u}\right) \d u\le C \int_s^t (1+\E y_u^2) \d u\le C\vert t-s\vert$. 
The inequality
\eqref{Novikov++} (with $p=2$)
implies directly that
 $$
 \E \left(
 \zeta_\zs{t}
 -
 \zeta_\zs{s}
  \right)^2
 \le C \vert t-s\vert
 $$ 
 for some constant $C>0$.
So,  the conclusion follows.
\endproof

\bibliographystyle{plain}
\bibliography{thesis}
\end{document}